\documentclass[12pt]{iopart}

\usepackage{amssymb}
\usepackage{graphicx}
\usepackage{color}
\usepackage{cite}

\def\be{\begin{equation}}
\def\ee{\end{equation}}
\def\gs{\mathrel{
   \rlap{\raise 0.511ex \hbox{$>$}}{\lower 0.511ex \hbox{$\sim$}}}}
\def\ls{\mathrel{
   \rlap{\raise 0.511ex \hbox{$<$}}{\lower 0.511ex \hbox{$\sim$}}}}
\newcommand{\obb}{0\mbox{$\nu\beta\beta$}}
\newcommand{\onbb}{neutrinoless double beta decay }
\newcommand{\meff}{\mbox{$|m_{ee}|$}}
\newcommand{\ba}{\begin{array}{c}}
\newcommand{\baz}{\begin{array}{cc}}
\newcommand{\bad}{\begin{array}{ccc}}
\newcommand{\bav}{\begin{array}{cccc}}
\newcommand{\baf}{\begin{array}{ccccc}}
\newcommand{\bas}{\begin{array}{cccccc}}
\newcommand{\bea}{\begin{equation} \begin{array}{c}}
\newcommand{\eea}{ \end{array} \end{equation}}
\newcommand{\ea}{\end{array}}
\newcommand{\D}{\displaystyle}

\begin{document}

\title[Neutrinoless Double Beta Decay]{Neutrinoless Double Beta Decay }

\author{Heinrich P\"as$^1$ and Werner Rodejohann$^2$}

\address{$^1$ Fakult\"at f\"ur Physik, Technische Universit\"at Dortmund, 
D-44221 Dortmund, Germany}
                
\address{$^2$ Max-Planck-Institut f\"ur Kernphysik, Postfach 103980, D-69029 Heidelberg, Germany
}
\eads{\mailto{heinrich.paes@tu-dortmund.de}, \mailto{werner.rodejohann@mpi-hd.mpg.de}}

\begin{abstract}
We review the potential to probe new physics with neutrinoless double beta decay 
$(A,Z) \to (A,Z+2) + 2 e^-$. Both the standard long-range light neutrino 
mechanism as well as short-range mechanisms mediated by heavy 
particles are discussed. We also stress aspects of the connection to 
lepton number violation at colliders and the implications for baryogenesis.

\end{abstract}

\submitto{\NJP}
\maketitle

\section{Introduction}
\label{sec:intro}

Neutrinoless double beta decay ($0\nu\beta\beta$) experiments are not simply neutrino mass experiments, 
but have a much more fundamental goal, namely the quest for lepton number violation (LNV). 
The basic decay mode is 
\begin{equation}\label{eq:main}
(A,Z) \to (A,Z+2) + 2 e^- \, ,
\end{equation}
which obviously violates electron lepton number $L_e$ by two units.
At present this endeavor is entering a particular exciting stage, with numerous
experiments operating or being under development, using different isotopes and experimental techniques (see Table \ref{tab:exp}). 
The previous best limit on the decay, set by the Heidelberg-Moscow experiment in 
2001~\cite{KlapdorKleingrothaus:2000sn}, has 
finally been improved from 2012 
on \cite{Auger:2012ar,Gando:2012zm,Agostini:2013mzu,Albert:2014awa}, 
and the limits will be further and further increased, with 
the potential of discovery always present. A large number of reviews 
has been written in the last few years 
\cite{Rodejohann:2011mu,GomezCadenas:2011it,Elliott:2012sp,Bilenky:2012qi,Rodejohann:2012xd,Deppisch:2012nb,Vergados:2012xy,Vogel:2012ja,Schwingenheuer:2012zs,Petcov:2013poa,Cremonesi:2013vla}, adding 
to the important earlier ones \cite{Haxton:1985am,Doi:1985dx,Vergados:2002pv,Avignone:2007fu,Simkovic:2007vu}, 
and emphasizing the importance of the decay and the strong interest of 
various communities. 

In this review we discuss the main physics potential and 
the conceptual implications that neutrinoless double beta decay brings along. 
We consider not only the standard three neutrino paradigm, but also 
different frameworks, including 
situations associated with heavy particle exchange, so-called short range mechanisms. 
Tests of such mechanisms are possible for instance in collider
experiments. In turn, observation of 
lepton number violation, either in $0\nu\beta\beta$ decay or at colliders, has important ramifications for baryogenesis, 
which we will outline as well. 

\begin{table}[pt]\hspace{-.cm}
{\small
{\hspace{-.cm}\begin{tabular}{c|c|ccc|c} \hline
Name & Isotope & \multicolumn{3}{c|}{source $=$ detector} 
& source $\neq$ detector \\ 
 &      & $\Delta E$ high & $\Delta E$ low 
& topology & topology\\ \hline
AMoRE & $^{100}$Mo & \checkmark & -- & -- & -- \\ 
CANDLES & $^{48}$Ca & -- & \checkmark & -- & -- \\ 
COBRA & $^{116}$Cd (and $^{130}$Te)  & -- &  -- & \checkmark &  -- \\ 
CUORE & $^{130}$Te  &\checkmark  &  -- &  -- &  -- \\
CUPID & $^{82}$Se / $^{100}$Mo / $^{116}$Cd / $^{130}$Te  &\checkmark  &  -- &  -- &  -- \\
DCBA/MTD  & $^{82}$Se / $^{150}$Nd  &  -- &  -- &  -- & \checkmark \\ 
EXO & $^{136}$Xe  &  -- &  -- & \checkmark &  -- \\ 
GERDA & $^{76}$Ge & \checkmark &  -- &  -- &  -- \\ 
KamLAND-Zen & $^{136}$Xe &  -- & \checkmark &  -- &  -- \\ 
LUCIFER & $^{82}$Se / $^{100}$Mo / $^{130}$Te  & 
\checkmark &  -- &  -- & -- \\ 
LUMINEU & $^{100}$Mo  & 
\checkmark &  -- &  -- & -- \\ 
 MAJORANA & $^{76}$Ge & \checkmark &  -- &  -- &  -- \\ 
MOON & $^{82}$Se / $^{100}$Mo / $^{150}$Nd 
 &  -- &  -- &  -- & \checkmark \\ 
NEXT & $^{136}$Xe &  -- &  -- & \checkmark &  -- \\ 
SNO+ & $^{130}$Te  &  -- & \checkmark &  -- &  -- \\ 
SuperNEMO& $^{82}$Se / $^{150}$Nd  &  -- &  -- &  -- & \checkmark \\ 
XMASS & $^{136}$Xe  &  -- & \checkmark &  -- &  -- \\
\hline
\end{tabular}}
}
\caption{\label{tab:exp}
Overview of present and future $0\nu\beta\beta$ decay experiments, their energy resolution and 
sensitivity to event topology.}
\end{table}

Why is it important to look for lepton number violation? One could give several reasons, 
for instance: 
\begin{itemize}
\item lepton number 
(as well as baryon number) is only an accidentally conserved global
symmetry in the Standard Model\footnote{Though not really connected 
to double beta decay or Majorana neutrinos that require  
lepton number violation by two units, one should note that even 
within the Standard Model lepton number is actually not conserved: chiral anomalies related 
to instanton tunneling break global lepton and baryon number by three
units each.}, 
and its conservation in extended theories seems very unlikely. Indeed, the lowest  
higher dimensional operator one can write down, 
${\cal L} = 1/\Lambda \, (\Phi L) \, (\Phi L)$, 
immediately violates lepton number and generates neutrino mass. In this language, neutrino 
mass and lepton number violation are the leading order new physics effects that one might expect to 
appear, as all other operators are suppressed by additional powers of the cut-off scale 
$\Lambda$. As neutrino mass has been observed in the form of neutrino oscillations, 
hopes are high that lepton number violation is present as well; 
\item 
the Universe contains more matter than antimatter. In order to generate this baryon asymmetry of the Universe,
baryon number conservation has to be violated. Unless nature treats baryon and lepton number in a 
completely different manner, also lepton number violation can be expected; 
\item in Grand Unified Theories lepton and baryon number are often connected, based on the 
fact that their difference can be gauged in an anomaly-free way when right-handed neutrinos 
are introduced. Thus baryon number violation typically 
implies lepton number violation. Moreover, GUTs usually implement a seesaw mechanism and thus 
Majorana neutrinos, leading eventually to $0\nu\beta\beta$ decay; 
\item almost all mechanisms that generate and suppress neutrino masses
  result in  Majorana 
neutrinos and thus eventually induce $0\nu\beta\beta$ decay; 
\item all theories beyond the Standard Model that violate lepton number by one or two 
units lead to neutrinoless double beta decay. Those include supersymmetric theories 
with $R$-parity violation, left-right symmetry theories, models with spontaneously broken 
lepton number, etc\footnote{We note here that if lepton number is
  violated not by two units but by three or more, there will be no neutrinoless
  double beta decay, but rather processes with $\Delta
  L=3,4,\ldots$ One explicit example is ``neutrinoless quadruple beta
  decay'' presented in Ref.\ \cite{Heeck:2013rpa}.}; 
\item in general, global symmetries are not expected to be conserved in quantum gravity  
theories. One could thus gauge lepton number, and in order to avoid long range forces one 
would need to break the gauge symmetry, leading again typically to lepton number violation. 
\end{itemize}
All in all, lepton number is not expected to be conserved, and the observation of 
lepton number violation would be as important as 
baryon number violation, e.g.\ proton decay. 
The decay width of double beta decay for a single operator inducing the decay can always be written as 
\be \label{eq:gamma}
\Gamma = 
 G(Q,Z)\, \left|{\cal M} \, \varepsilon \right|^2 ,
\ee
where $G(Q,Z)$ is a calculable phase space factor typically scaling with the endpoint energy 
as $Q^5$ and ${\cal M}$ is the nuclear matrix element, which is notoriously difficult to calculate.
The particle physics parameter $\varepsilon$, which depends on 
particle masses, mixing parameters etc., is most important 
from the point of view of this review. Note that more than one mechanism can contribute, 
hence the amplitude of the decay can actually be 
\be \label{eq:ampl}
{\cal A} = \sum\limits_x {\cal M}_x \, \varepsilon_x \, ,
\ee
i.e.\ a sum over different mechanisms, which can potentially 
interfere with each other.  
\\

The review is organized as follows: in Section \ref{sec:nu} we summarize double 
beta decay mediated by light massive Majorana neutrinos while  
Section \ref{sec:short} deals with alternative and 
short-range mechanisms, including potential tests. The connection between $0\nu\beta\beta$ decay,
lepton number violation at colliders and baryogenesis is discussed in Section 
\ref{sec:YB}, before we conclude\footnote{Topics that are not covered in this review are the 
experimental and nuclear physics aspects, where the interested reader should 
consult e.g.\ the review articles 
\cite{GomezCadenas:2011it,Schwingenheuer:2012zs,Cremonesi:2013vla} and \cite{Vogel:2012ja}, 
respectively.}  in Section \ref{sec:conc}.

\section{Neutrinoless Double Beta Decay and Neutrino Masses}
\label{sec:nu}

\begin{figure}[t]
\begin{center}
\includegraphics[width=.45\textwidth,angle=0]{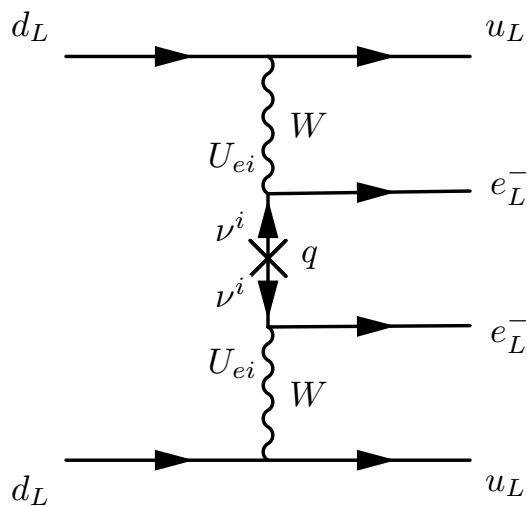}
\caption{\label{fig:SI}Feynman diagram for the standard interpretation (mass mechanism) of
  neutrinoless double beta decay.}
\end{center}
\end{figure}

We begin with the arguably best motivated possibility for the decay, the 
``standard interpretation'' or "mass mechanism", namely that the light massive neutrinos that we observe 
to oscillate in terrestrial experiments mediate double beta decay. In this case, searches 
for the process are searches for neutrino mass, complementing the other approaches to determine neutrino 
masses. Those approaches include direct searches in classical Kurie-plot experiments like the upcoming 
KATRIN \cite{Osipowicz:2001sq}, Project 8 \cite{Monreal:2009za}, ECHo \cite{Blaum:2013pfu} 
or MARE \cite{Monfardini:2005dk} experiments, and cosmological observations, see 
\cite{Lesgourgues:2014zoa} for a review in this Focus Issue. 
Cosmology probes the sum of neutrino masses, 
\be \label{eq:sigma}
\Sigma = \sum m_i\, , 
\ee
Kurie-plot experiments test the incoherent sum 
\be \label{eq:mbeta}
m_\beta = \sqrt{\sum |U_{ei}|^2 \, m_i^2} \, , 
\ee
whereas \onbb in the standard interpretation tests the quantity (see Fig.\ \ref{fig:SI})
\be \label{eq:meff}
\meff = \left| \sum U_{ei}^2 \, m_i \right| , 
\ee
which is usually called the effective mass and coincides with the $ee$ element of the
neutrino mass matrix in flavor space. 

\begin{figure}[t]
\begin{center}
\includegraphics[width=.45\textwidth,angle=270]{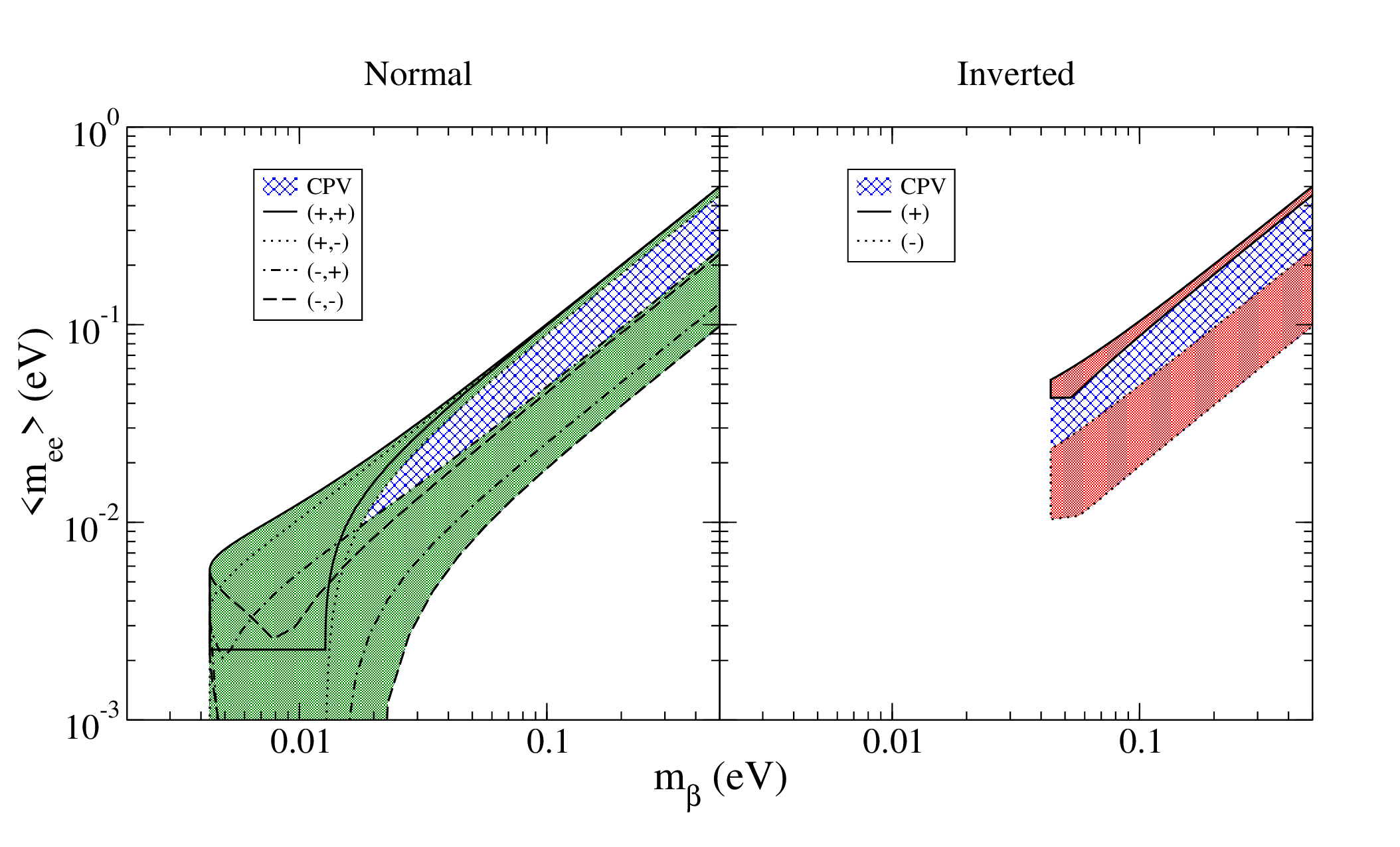}
\includegraphics[width=.45\textwidth,angle=270]{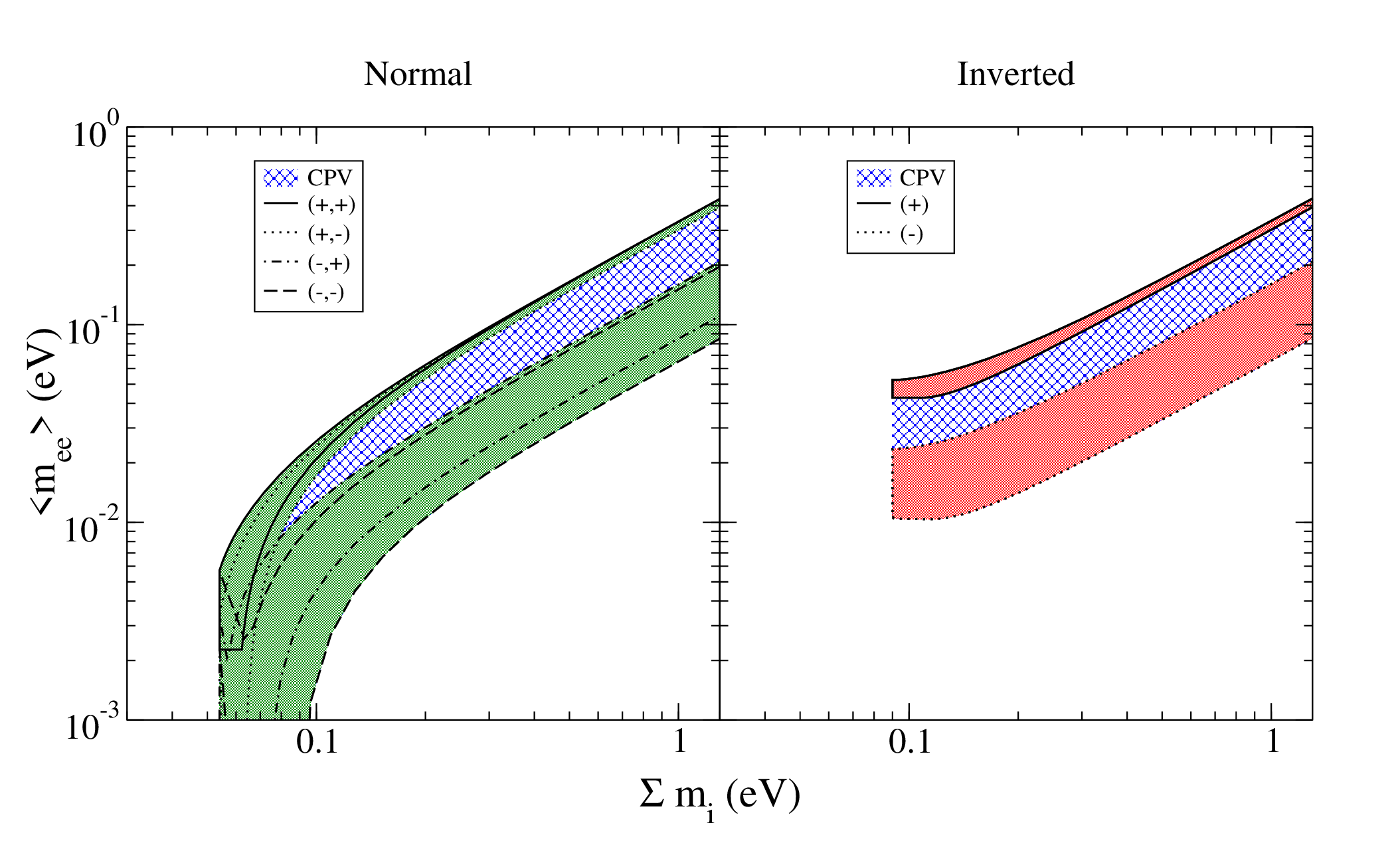}
\caption{\label{fig:usual}Neutrino mass observables within the standard 
three neutrino paradigm. The upper plot shows the effective mass
against the kinematical 
neutrino mass accessible in Kurie-plot experiments, the lower plot depicts the effective mass 
against the sum of masses accessible in cosmological measurements. The values for relative 
signs of the mass eigenvalues, and the areas which only can be realized for non-trivial CP phases 
are indicated.}
\end{center}
\end{figure}

Here $m_i$ are the neutrino masses, and $U_{ei}$ are elements of the leptonic mixing, 
or PMNS, matrix that is usually parametrized as 
\be \label{eq:U}
U = \left( \bad 
c_{12}   c_{13} 
& s_{12}  c_{13} 
& s_{13}  e^{-i \delta}  \\ 
-s_{12}  c_{23} 
- c_{12}  s_{23} \, 
s_{13}   e^{i \delta} 
& c_{12}  c_{23} - 
s_{12}  s_{23}  s_{13} 
\, e^{i \delta} 
& s_{23}   c_{13}  \\ 
s_{12}    s_{23} - c_{12} 
 c_{23}  s_{13}  e^{i \delta} & 
- c_{12}  s_{23} 
- s_{12}  c_{23} \, 
s_{13}  e^{i \delta} 
& c_{23}   c_{13}  
\ea   
\right) P \,,
\ee
where $s_{ij} = \sin \theta_{ij}$, $c_{ij} = \cos \theta_{ij}$ and
$\delta$ is the ``Dirac phase'' responsible for CP violation in
neutrino oscillation experiments. The diagonal phase matrix 
$P= {\rm diag}(1,e^{i \alpha}, e^{i (\beta + \delta)})$ contains the two 
Majorana phases $\alpha$ and $\beta$, 
which are associated with the Majorana nature of neutrinos and thus only 
show up in lepton number violating processes (a review on properties of 
Majorana particles can be found in \cite{Akhmedov:2014kxa}). 
For three neutrinos we have therefore 9 physical parameters, three
masses $m_{1,2,3}$, three mixing angles $\theta_{12}, \theta_{13},
\theta_{23}$ and three phases $\delta, \alpha, \beta$. The effective mass depends thus on 
7 out of those 9 physical neutrino parameters: 
\be
\meff = f(\theta_{12}, \theta_{13}, \alpha, \beta, m_1, m_2, m_3)\,.
\ee
Of these seven parameters, we currently do not know the phases and the 
lightest mass, where in addition 
the mass ordering is unknown, i.e.\ it could be either $m_3 > m_2 > m_1$ (normal ordering) 
or $m_2 > m_1 > m_3$ (inverted ordering). Global fits of all available neutrino data can be 
found in Refs.\ \cite{Forero:2014bxa,Capozzi:2013csa,Gonzalez-Garcia:2014bfa}. 
One can then use Eqs.\ (\ref{eq:sigma}, \ref{eq:mbeta}, \ref{eq:meff}) to plot the 
three neutrino mass observables against each other \cite{Fogli:2004as}, 
see Fig.\ \ref{fig:usual}, and interpret potential current and future experimental results.

For instance, in case one finds positive results for $m_\beta$ and $\meff$ in any 
of the green or red areas in the upper plot of Fig.\ \ref{fig:usual}, then this would 
be a convincing confirmation 
of the paradigm that there are three massive Majorana neutrinos 
mixing among each other. Even more spectacular would be if inconsistencies arise, e.g.\ 
a measurement of the effective mass that is incompatible with limits from 
KATRIN or cosmology. This would imply that something in our interpretation of double 
beta decay goes amiss, i.e.\ that another mechanism causes the decay. Therefore, the 
complementarity of the various approaches to determine the neutrino mass offers exciting possibilities, since different 
assumptions enter their interpretation. KATRIN-like experiments are essentially 
model-independent, as only bizarre things like tachyonic neutrinos could spoil the results, 
and moreover the interpretation is ``clean'' as beta decay is theoretically well under 
control. However, in terms of numbers the limits are and will be the weakest, and 
further improvement beyond 0.1 eV seems impossible. Cosmology yields  the best 
limits in terms of numbers, and can even contribute to the question of mass ordering. However, it suffers 
from difficult systematics and relies on model input, e.g.\ departures from simple 
$\Lambda$CDM models can weaken limits considerably. Double beta decay is the most fundamental 
approach as it is connected to lepton number violation, and can even say something about the 
mass ordering (see below). However, it is very model-dependent as many mechanisms
apart from the standard neutrino mass mechanism can mediate the decay. Furthermore, the process 
is theoretically ``dirty'', as nuclear matrix element introduce a sizable uncertainty. 
The pros and cons of the different approaches and their current as well as near and far 
future limits are summarized in Table \ref{tab:nu}.\\

\begin{table}[pt]
\begin{tabular}{ccccc||c|c} \hline
Method & observable & now [eV] & near [eV] & far [eV] & pro & con \\ \hline \hline
\raisebox{-1.5ex}{Kurie} &  \raisebox{-1.5ex}{$\sqrt{\sum |U_{ei}|^2 \, m_i^2}$} & \raisebox{-1.5ex}{$2.3$} & \raisebox{-1.5ex}{$0.2$} & \raisebox{-1.5ex}{$0.1$} & {\scriptsize model-indep.;} & {\scriptsize final?;} \\[-.325cm]  
       &            &     &      &       & {\scriptsize theo.\ clean} & {\scriptsize worst} \\ \hline 
\raisebox{-1.5ex}{Cosmo.} & \raisebox{-1.5ex}{$\sum m_i$} & \raisebox{-1.5ex}{$0.7$} & \raisebox{-1.5ex}{$0.3$} &\raisebox{-1.5ex}{ $0.05$} & {\scriptsize best;} & {\scriptsize systemat.;}  \\[-.325cm]  
       &            &     &      &       & {\scriptsize NH/IH} & {\scriptsize model-dep.} \\ \hline 
\raisebox{-1.5ex}{\obb} & \raisebox{-1.5ex}{$|\sum U_{ei}^2 m_i|$} & \raisebox{-1.5ex}{$0.3$} & \raisebox{-1.5ex}{$0.1$} & \raisebox{-1.5ex}{$0.05$} & {\scriptsize fundament.;} & {\scriptsize model-dep.;} 
\\[-.325cm]  
& & & & & {\scriptsize NH/IH} & {\scriptsize theo.\ dirty} \\ \hline \hline
\end{tabular}
\caption{\label{tab:nu}Summary of the main approaches to neutrino mass.}
\end{table}

\begin{figure}[t]
\begin{center}
\includegraphics[width=.45\textwidth,angle=270]{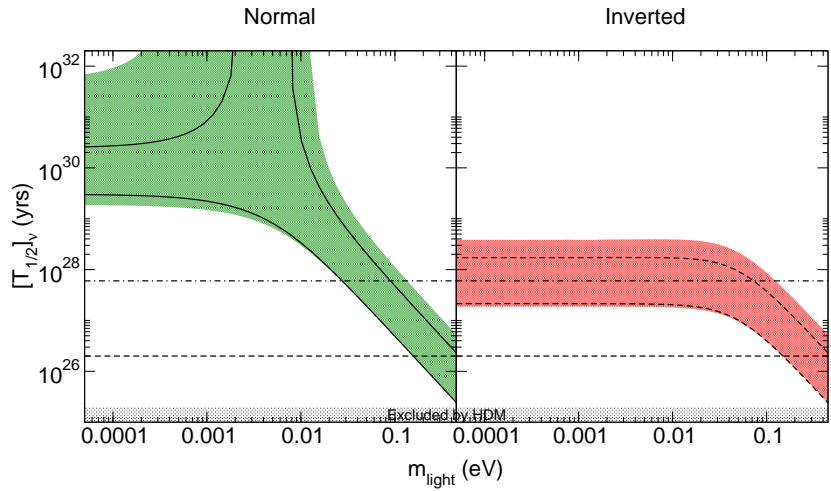}
\caption{\label{fig:lifetime}Example for typical half lives corresponding to $^{76}$Ge and a matrix 
element of ${\cal M}_{\rm Ge}=4.6$.}
\end{center}
\end{figure}

It is important to note that for the normal mass ordering the effective mass can vanish,  
whereas for the inverted ordering the effective mass cannot 
vanish \cite{Pascoli:2002xq}. Hence, the lifetime in this latter case is necessarily finite, though 
of course an experimental challenge. The lower limit is given by  
\be
\meff_{\rm min}^{\rm IH} = \sqrt{\Delta m^2_{31}} \,c_{13}^2 \, (1 - 2 \sin^2 \theta_{12}) \simeq 
(0.01 \ldots 0.02)~{\rm eV}\, , 
\ee
corresponding to half-lives around $10^{27}$ yrs, see Fig.\ \ref{fig:lifetime}.  
This minimal value depends rather strongly on the solar neutrino mixing angle $\theta_{12}$. 
Hence, a more precise determination of $\theta_{12}$ in future oscillation experiments 
would be rather welcome \cite{Dueck:2011hu}. 
Within the well-motivated three Majorana neutrino paradigm the upper
and lower value of the effective mass in the inverted
ordering are the natural medium-term goal for neutrinoless double beta decay
searches. 
In case the mass ordering turns out to be normal, this
motivation is lost. However, the value of neutrino mass remains
unknown, and consistency checks with cosmological or Kurie-plot limits
are necessary. Moreover, as argued in the introduction, 
the highly important search for lepton number violation needs to be pursued further.

What is the current limit on the effective mass? To answer this question, a comparison of 
different isotopes and matrix elements is necessary. One of the most competitive lifetime 
limits is set by GERDA~\cite{Agostini:2013mzu}, 
$T_{1/2}^{\rm Ge} > 2.1 \cdot 10^{25}$ yrs, or, combined with earlier
Germanium experiments \cite{KlapdorKleingrothaus:2000sn,Aalseth:2002rf}, 
$T_{1/2}^{\rm Ge} > 3.0 \cdot 10^{25}$ yrs. A similarly strong limit is obtained by the 
KamLAND-Zen experiment \cite{Gando:2012zm}, namely $T_{1/2}^{\rm Xe} > 2.6 \cdot 10^{25}$ yrs. 
Using Eq.\ (\ref{eq:gamma}), one finds that experiments using $^{136}$Xe give a better limit 
than experiments with $^{76}$Ge if their lifetime limit fulfills the condition: 
\be 
T^{\rm Xe}_{1/2} > T^{\rm Ge}_{1/2}\, \frac{G_{\rm Ge}}{G_{\rm Xe}} \, 
\left|\frac{{\cal M}_{\rm
Ge}}{{\cal M}_{\rm Xe}}\right|^2 \, {\rm yrs}.
\ee
Using the phase space factors of Refs.\ \cite{Kotila:2012zza,Mirea:2014dza}, and the matrix 
elements of various groups, the limits on the effective mass in 
Table \ref{tab:limits} are obtained, adapted from \cite{Dev:2013vxa}. Some matrix 
element approaches have a better limit from Germanium, others from Xenon. Taking correctly 
the conservative values, both isotopes give essentially the same limit
of\footnote{See also Ref.\ \cite{Guzowski:2015saa} for an approach to combine different
  experiments in a statistical manner.} 
\be
\meff \ls 0.3\,{\rm eV}.
\ee
Future improvement of this limit goes with the square root of lifetime limits. \\

\begin{table}[t]
\begin{center}
\begin{tabular}{c|c|c|c}\hline
 NME & \multicolumn{2}{|c}{ $^{76}$Ge} &
\multicolumn{1}{|c}{ $^{136}$Xe} \\ \cline{2-4}
&   GERDA &   combined  &  KamLAND-Zen  \\ \hline
{\bf  EDF(U)} \cite{Rodriguez:2010mn} & {\bf  0.32} & {\bf  0.27}  & {\bf  0.13}  \\ 
{\bf  ISM(U)} \cite{Menendez:2008jp}&  {\bf  0.52} & {\bf  0.44 } & {\bf  0.24}  \\ 
{\bf  IBM-2}\cite{Barea:2013bz} &  {\bf  0.27} & {\bf  0.23}  & {\bf 
  0.16} \\ 
{\bf  pnQRPA(U)} \cite{Suhonen:2010zzc} &  {\bf  0.28} & {\bf  0.24}  & {\bf  0.17}  \\
{\bf  SRQRPA-B} \cite{Meroni:2012qf}& {\bf  0.25} & {\bf  0.21}  & {\bf  0.15} \\ 
{\bf  SRQRPA-A} \cite{Meroni:2012qf} & {\bf  0.31} & {\bf  0.26}  & {\bf  0.23}  \\ 
{\it  QRPA-A} \cite{Simkovic:2013qiy}& {\it  0.28} & {\it  0.24}  & {\it  0.25}  \\ 
{\it  SkM-HFB-QRPA} \cite{Mustonen:2013zu} &  {\it  0.29} & {\it  0.24}  & {\it  0.28}  \\ 
\hline
\end{tabular}
\caption{\label{tab:limits} Limits on the effective mass $m_{ee}$ (in eV) 
from Germanium and Xenon experiments 
and different matrix element calculations. The calculations listed in bold face yield a better limit 
for  $^{136}$Xe, the ones in italic give a better limit 
for  $^{76}$Ge. Adapted from \cite{Dev:2013vxa}.}
\end{center}
\end{table}

So far the effective mass has simply been used as a phenomenological parameter. 
Of course, in case one has a model at hand, one can predict \meff~to some extent.   
One example are popular flavor symmetry models to explain the peculiar features 
of lepton mixing \cite{Altarelli:2010gt,Ishimori:2010au}. While the neutrino mass itself 
cannot be predicted in this framework, relations between neutrino masses are possible to predict, so-called 
neutrino mass sum-rules such as $\tilde m_1 + \tilde m_2 = \tilde m_3$. Here the masses are understood to 
be complex, i.e.\ including the Majorana phases. These relations exclude some possible combinations of 
masses and phases, and thus only certain areas in parameter space are possible, 
which allows to rule out certain models. Many sum-rule examples have been discussed in 
the literature \cite{Barry:2010yk,Dorame:2011eb,Dorame:2012zv,King:2013psa}. 
Even more predictive are some
Grand Unified Theories, where the Yukawa matrices of all fermions are 
related and fitting the constrained matrices to the observed mass and mixing parameters 
allows to predict unknown parameters such as \meff, see \cite{Dueck:2013gca}. \\

While the three neutrino paradigm is very attractive and robust, there are longstanding 
hints that light sterile neutrinos with mass around an eV and mixing around 10\,\% exist, 
see Ref.\ \cite{Abazajian:2012ys} for a review of the various hints and 
ongoing as well as future tests. Such a fourth neutrino would modify all neutrino 
mass observables, in particular the effective mass: 
\be 
\meff = | \underbrace{|U_{e1}|^2  \, m_1 + |U_{e2}|^2  \, m_2 \, e^{2 i
\alpha} + |U_{e3}^2|  \, m_3 \, e^{2 i \beta} 
}_{\D m_{ee}^{\rm act}} + 
\underbrace{|U_{e4}|^2  \, m_4 \, e^{2 i \gamma}}_{\D m_{ee}^{\rm st}} | \, , 
\ee
where $\gamma$ is an additional Majorana phase and $m_{ee}^{\rm act}$ the three neutrino 
contribution discussed so far. The sterile contribution $\meff^{\rm st}$ 
to \obb~(assuming a 1+3 scenario) generates typical values of the same order as 
$m_{ee}^{\rm act}$ for the inverted ordering: 
\be \label{eq:sterile}
\meff^{\rm st} \simeq \sqrt{\Delta m^2_{\rm st}} \, |U_{e4}|^2  \left\{ 
\ba 
\gg \meff_{\rm NH}^{\rm act} \\
\simeq \meff_{\rm IH}^{\rm act}
\ea 
\right. .
\ee
Thus, in contrast to the three-generation case, for a normal mass ordering of the active neutrinos the effective mass 
cannot vanish anymore, whereas for an inverted ordering of the active neutrinos the 
effective mass can vanish now \cite{Barry:2011wb,Giunti:2012tn,Girardi:2013zra,Giunti:2015kza}. 
The phenomenology has completely turned around! This  
demonstrates that any physics output of neutrinoless double beta decay depends dramatically 
on the assumptions. 

We are thus naturally lead to discuss alternative mechanisms of double beta decay, 
to be addressed in the following Section.

\section{Neutrinoless Double Beta Decay and Short-Range Mechanisms}
\label{sec:short}

\begin{figure}
\centering
\includegraphics[clip,width=0.4\textwidth]{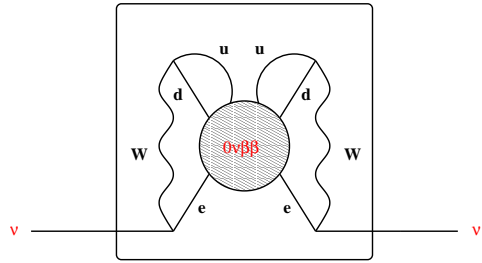}
\caption{\label{fig:bbth}Black box theorem depicted as a Feynman diagram:
neutrinoless double beta decay always induces a neutrino Majorana mass
(from \protect{\cite{Deppisch:2012nb}}).
}
\end{figure}

Apart from the standard interpretation
where a massive Majorana neutrino is being exchanged between Standard Model (SM) $V-A$ vertices, in principle any operator violating lepton number by two units and transforming two neutrons into
two protons, two electrons and nothing else will induce the decay.
This does not mean, however, that neutrinoless double beta decay and
the question whether the neutrino possesses a Majorana
mass are totally decoupled: 
the observation of neutrinoless double beta decay demonstrates that
lepton number is violated by two units. Such lepton number violation implies that
neutrinos have to be Majorana particles. That the two are inseparably
connected can be proven by what is known as the black box theorem
\cite{Schechter:1981bd, Nieves:1984sn, Takasugi:1984xr,Hirsch:1998mc,Hirsch:2006yk}. Graphically
the theorem can be depicted as shown in Fig.~\ref{fig:bbth}: If double
beta decay has been seen, a Majorana neutrino mass 
term is generated at 4-loop order, even if the underlying 
particle physics model does not contain a tree-level neutrino mass. 
Of course this contribution to the neutrino mass is rather small
\cite{Duerr:2011zd}, namely of order $G_F^4 /(16\pi^2)^4 m_{u,d,e}^5 \sim
10^{-25}$ eV, and thus clearly neither the dominant contribution to neutrinoless
double beta decay nor to neutrino mass itself. 
Note that this 4-loop contribution is only the minimal,
guaranteed connection between neutrino mass and double beta
decay arising in any scenario with $\Delta L_e=2$ LNV. 
Explicit models leading to $0\nu\beta\beta$ can generate
neutrino mass at tree, 1-, 2- or 3-loop level. Depending on the model, 
the neutrino masses generated in this way can lead to a 
comparable, sub-dominant or dominant neutrino contribution to the decay, and/or to a 
main, sub-leading or negligible contribution to neutrino mass. For a
comparative analysis of all scalar-mediated models based on the SM gauge group
see~\cite{Helo:2015fba}.

 \begin{figure}[!t]
\centering
\includegraphics[width=4in]{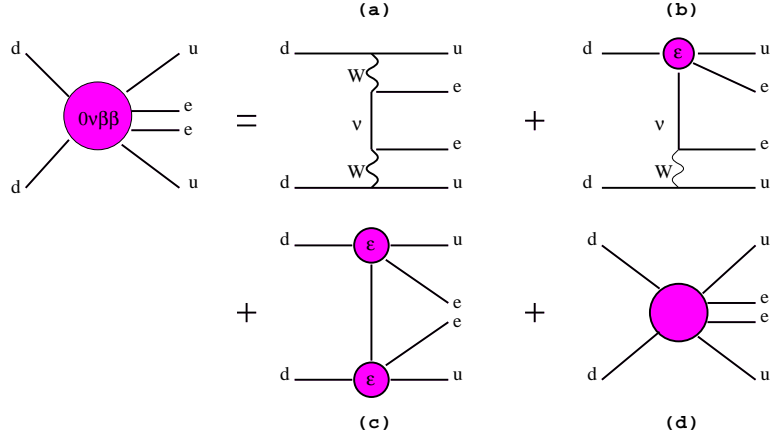}
\caption{Mechanisms for neutrinoless double beta decay:  the most general effective operator triggering the decay can be decomposed
into diagrams with SM vertices and effective vertices being point-like
at the nuclear Fermi momentum scale ${\cal O}(100)~{\rm MeV}$ 
(from \protect{\cite{Pas:1999fc}}).}
\label{general}
\end{figure}

The 
most general operator inducing the decay can be parametrized
in terms of effective couplings $\varepsilon$ (see Fig.~\ref{general}) \cite{Pas:1999fc,Pas:2000vn}.
The diagram depicts the standard interpretation exchange of a light Majorana neutrino between two SM vertices
(contribution a)), the exchange of
a light Majorana neutrino between an SM vertex and an effective operator which is point-like at the nuclear Fermi momentum scale
(the inverse size of the nucleon)
${\cal O}(100)~{\rm MeV}$ (contribution b)),
contribution c), which contains two non-SM vertices and can be neglected when compared to contribution b),
and a short-range contribution triggered by a single dimension 9 operator being point-like at
the Fermi momentum scale (contribution d)). 

We can estimate the energy scale of short-range diagrams which can
lead to comparable double beta decay lifetimes compared with the standard interpretation. 
The standard diagram
discussed in Section \ref{sec:nu} has an amplitude of order $G_F^2 \,
|m_{ee}|/q^2$. If the decay is mediated by  particles heavier than the
characteristic momentum scale of $q \simeq 100$ MeV, then the corresponding
amplitude is $c/M^5$, where $M$ is the mass of those particles and $c$
a combination of flavor and possible gauge coupling parameters. Hence,
for $c$ of order one and $M$ of order TeV this amplitude equals the
current limit on the standard amplitude (ignoring here a small
suppression of the nuclear matrix elements for short-range diagrams): 
\begin{equation}\label{eq:eVTeV}
T_{1/2}^{0\nu\beta\beta} (m_\nu = 1\, {\rm eV}) \simeq 
T_{1/2}^{0\nu\beta\beta} (M = 1\, {\rm TeV})\,.
\end{equation}
We thus can test short-range diagrams for double beta decay
with the LHC or lepton flavor violation experiments, which are also 
sensitive to the TeV scale. 

The most general decay rate contains all combinations of leptonic and hadronic currents induced by the operators
\begin{equation}
{\cal O}_{V\mp A} = \gamma^{\mu}(1\mp \gamma_5)\,, \,\,
{\cal O}_{S \mp P} = (1 \mp \gamma_5)\,,    \,\,    
{\cal O}_{T_{L/R}} = \frac{i}{2}[\gamma_{\mu},\gamma_{\nu}](1\mp \gamma_5)\,, 
\label{ops}
\end{equation}
allowed by Lorentz invariance.

Examples for contribution b) are the leptoquark and $R$-parity violating SUSY accompanied decay modes,
examples for contribution d) are decay modes where only SUSY particles or heavy neutrinos and gauge bosons
in left-right-symmetric models are exchanged between the decaying nucleons. 
Present experiments have a sensitivity to the effective couplings of 
\begin{equation}
\varepsilon < {\rm few} \cdot (10^{-7}-10^{-10}) \,.
\end{equation}
For a more detailed, recent overview on this approach to 
double beta decay see~\cite{Deppisch:2012nb}.  

As has been pointed out above for the
$d=9$ operator triggering the contribution d) it can be
estimated 
that an observation of $0\nu\beta\beta$ decay with present-day
experiments would involve TeV scale particles and thus would offer good chances to see new physics associated with LNV at the
LHC. A crucial prerequisite for such a conclusion is of course a possibility to discriminate among the various mechanisms which may
be responsible for the decay. This is a difficult task but may be possible at least for some of the mechanisms by 
observing neutrinoless double beta decay in multiple isotopes 
\cite{Deppisch:2006hb,Gehman:2007qg,Faessler:2011qw,Meroni:2012qf}.   
or by measuring the decay distribution, for example in the SuperNEMO
experiment \cite{Arnold:2010tu}. Another possibility to discriminate between various short-range contributions to neutrinoless
double beta decay at the LHC itself is to identify the invariant mass peaks of particles produced resonantly in the intermediate state or
to analyze the charge asymmetry between final states involving particles and/or anti-particles
\cite{Helo:2013dla,Helo:2013ika}.

\subsection{Left-Right Symmetry}
\label{sec:lrsymmetry}

\begin{figure}[t]
\begin{center}
\includegraphics[width=.215\textwidth,angle=0]{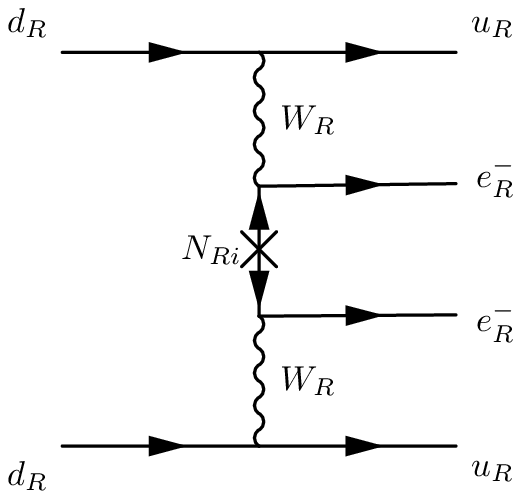}
\includegraphics[width=.215\textwidth,angle=0]{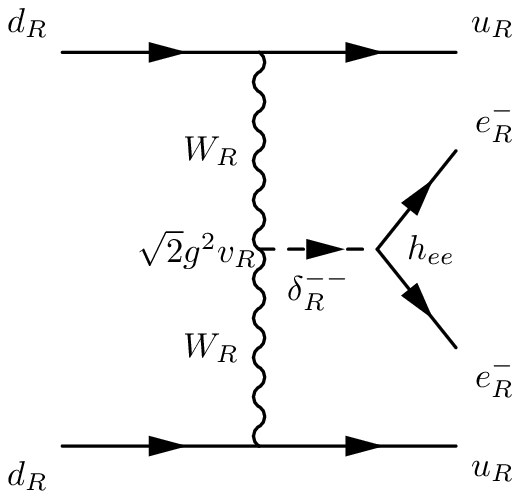}
\includegraphics[width=.215\textwidth,angle=0]{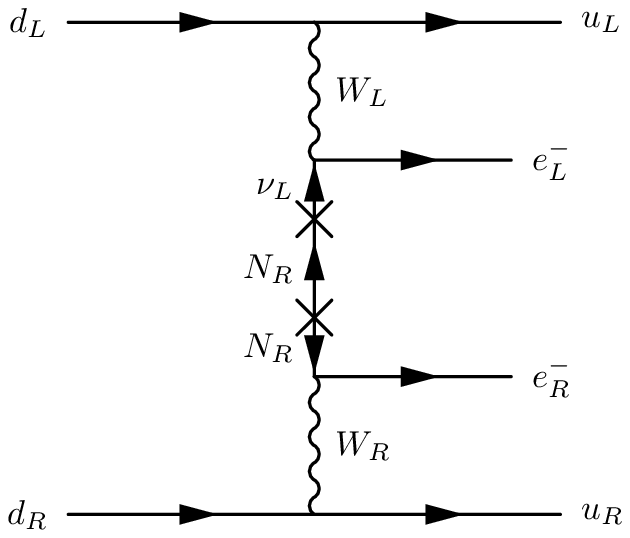}
\includegraphics[width=.215\textwidth,angle=0]{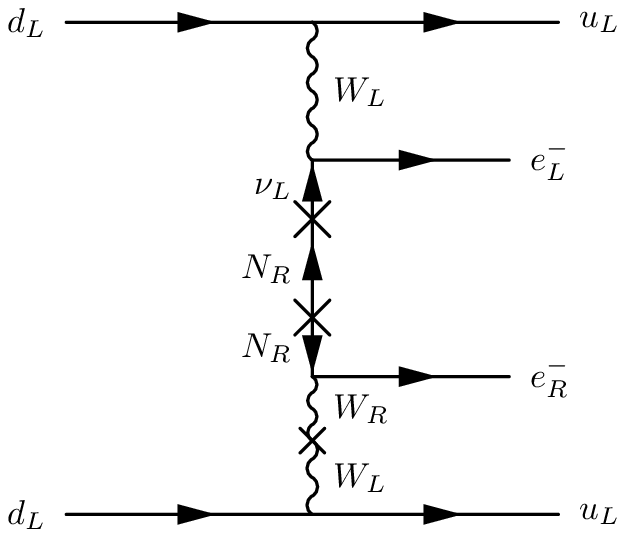}
\caption{\label{fig:LR}Feynman diagrams for the leading diagrams of
  neutrinoless double beta decay in left-right symmetric theories: 
  heavy neutrino exchange with right-handed currents, triplet
  exchange, the $\lambda$- and the $\eta$-diagram (from left to right).}
\end{center}
\end{figure}

In
Left-Right Symmetric Models the Standard Model gauge symmetry
is extended to the group SU(2)$_L~\otimes$ SU(2)$_R~\otimes$
U(1)$_{B-L}$. Right-handed neutrinos are a necessary ingredient to
realize this extended symmetry and are included in an SU(2)$_R$
doublet. A generation of leptons is assigned to the
multiplets $L_i = (\nu_i, l_i)$ with the quantum numbers $Q_{L_L} =
(1/2, 0, -1)$ and $Q_{L_R} = (0, 1/2, -1)$ under SU(2)$_L~\otimes$
SU(2)$_R~\otimes$ U(1)$_{B-L}$. The Higgs sector contains a bidoublet
$\phi$ and two triplets $\Delta_L$ and $\Delta_R$. The VEV $v_R$ of
the neutral component of $\Delta_R$ breaks SU(2)$_R~\otimes$
U(1)$_{B-L}$ to U(1)$_Y$ and generates masses for the right-handed
$W_R$ and $Z_R$ gauge bosons, and the heavy neutrinos. Since
right-handed currents and particles have not been observed, $v_R$ has
to be sufficiently large. Neutrino masses are then generated within a
type-I+II seesaw, 
\begin{equation}
m_\nu = m_L - m_D M_R^{-1} m_D^T\,,
\end{equation} 
where $m_L = f v_L$ and $M_R = f v_R$ are the VEVs of the triplets. 
Within left-right symmetric models several diagrams mediating double beta decay
exist, see Fig.\ \ref{fig:LR}. 

The right-handed neutrinos and $W_R$ bosons can mediate the
right-handed analogue of the standard mechanism discussed above
\cite{Doi:1985dx, Muto:1989cd,Hirsch:1996qw}. 
As the particles exchanged are much heavier than the nuclear Fermi
momentum, 
  this is a realization of the short-range operator. The now heavy
  neutrino mass will appear in the denominator of the amplitude 
instead of the  numerator.  
The effective coupling is denoted $\varepsilon_3^{RRz}$. Assuming
manifest left-right symmetry, i.e.\ identical gauge couplings, 
in terms of the left-right-symmetric model parameters it is given by
\begin{equation}
\label{eq:epsilonN}
	\varepsilon_3^{RRz} = \sum_{i=1}^3 V_{ei}^2 	
	\frac{m_p}{m_{N_i}}\frac{m_{W_L}^4}{m_{W_R}^4} \,,
\end{equation}
where $V$ denotes the matrix describing the
mixing among the heavy right-handed neutrinos. 
Searches for $0\nu\beta\beta$ yield the limit $|\varepsilon_3^{RRz}|
\ls 1 \cdot 10^{-8}$. 

Moreover, 
because of the presence of right-handed currents, the exchange of light neutrinos does not necessarily require a chirality violating mass insertion. 
The coupling parameters of the corresponding effective long-range operators can be written as
\begin{equation}
\label{eq:epsilonLR}
	\varepsilon_{V+A}^{V+A} = \sum_{i=1}^3 U_{ei} S_{ei} 	
	\frac{m_{W_L}^2}{m_{W_R}^2}\,, \qquad
	\varepsilon_{V-A}^{V+A} = \sum_{i=1}^3 U_{ei} S_{ei} 	
	\tan\zeta\,,
\end{equation}
with the current experimental limits $|\varepsilon_{V+A}^{V+A}| \ls 5 \cdot 10^{-7}$ and $|\varepsilon_{V-A}^{V+A}| \ls 3 \cdot 10^{-9}$, respectively, and where
$S$ describes the mixing between left- and right-handed neutrinos. 
 The diagram governed by $\varepsilon_{V+A}^{V+A} $ is often called
the $\lambda$-diagram, the one governed by $\varepsilon_{V-A}^{V+A}$ the
$\eta$-diagram.  While the mixing $S$ is small in the simplest seesaw
scenarios, one can easily arrange for large left-right (or equivalently light-heavy) 
mixing. In this case both diagrams can be expected to dominate over the heavy
neutrino exchange diagram with right-handed currents \cite{Barry:2013xxa,Dev:2014xea}. 
Analyses of the type-I seesaw
mechanism with sizable light-heavy mixing can be found in
\cite{Ibarra:2010xw,Mitra:2011qr}. 

Finally, there exists a contribution from the exchange of a right-handed doubly-charged
Higgs triplet $\Delta_R$, which has the same effective operator
structure as heavy neutrino exchange. The effective short-range
coupling strength is here given as 
\begin{equation}
\label{eq:epsilonDelta}
	\varepsilon_3^{RRz} = \sum_{i=1}^3 V_{ei}^2 		
	\frac{m_{N_i}m_p}{m^2_{\Delta_R}}\frac{m_{W_L}^4}{m_{W_R}^4} \ls 1.1 \cdot 10^{-8},
\end{equation}
%
Since the Higgs triplet can mediate $\mu \to 3e$ at
tree level there are strong constraints on this diagram by lepton
flavor violation bounds \cite{Barry:2013xxa}. 

A particularly predictive case occurs if type-II seesaw dominance
holds, i.e.\ if the neutrino mass matrix is generated by the SU(2)$_L$
triplet term $m_L$. Due to the discrete left-right symmetry this term
is directly proportional to the heavy neutrino mass matrix, hence $V$ in
Eq.\ (\ref{eq:epsilonDelta})  equals the PMNS matrix $U$ and $m_i
\propto M_i$. It follows \cite{Tello:2010am} that
typically for a normal mass ordering the
lifetime of double beta decay is finite while for an inverted mass
ordering it can be infinite due to possible cancellations. Just as for
the case of light sterile neutrinos (see Eq.\ (\ref{eq:sterile})) 
the standard phenomenology has turned around.

\begin{figure}[t]
\begin{center}
\includegraphics[width=.4215\textwidth,angle=0]{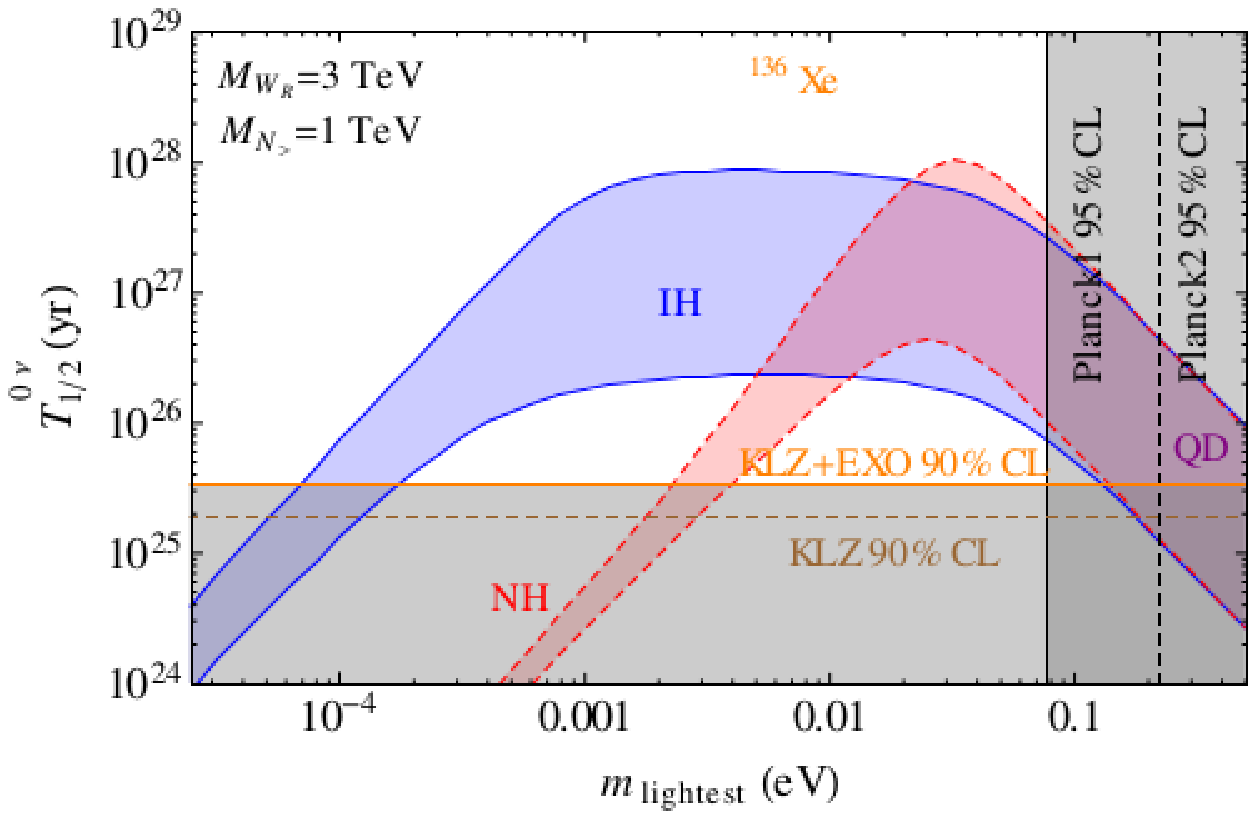}
\includegraphics[width=.4215\textwidth,angle=0]{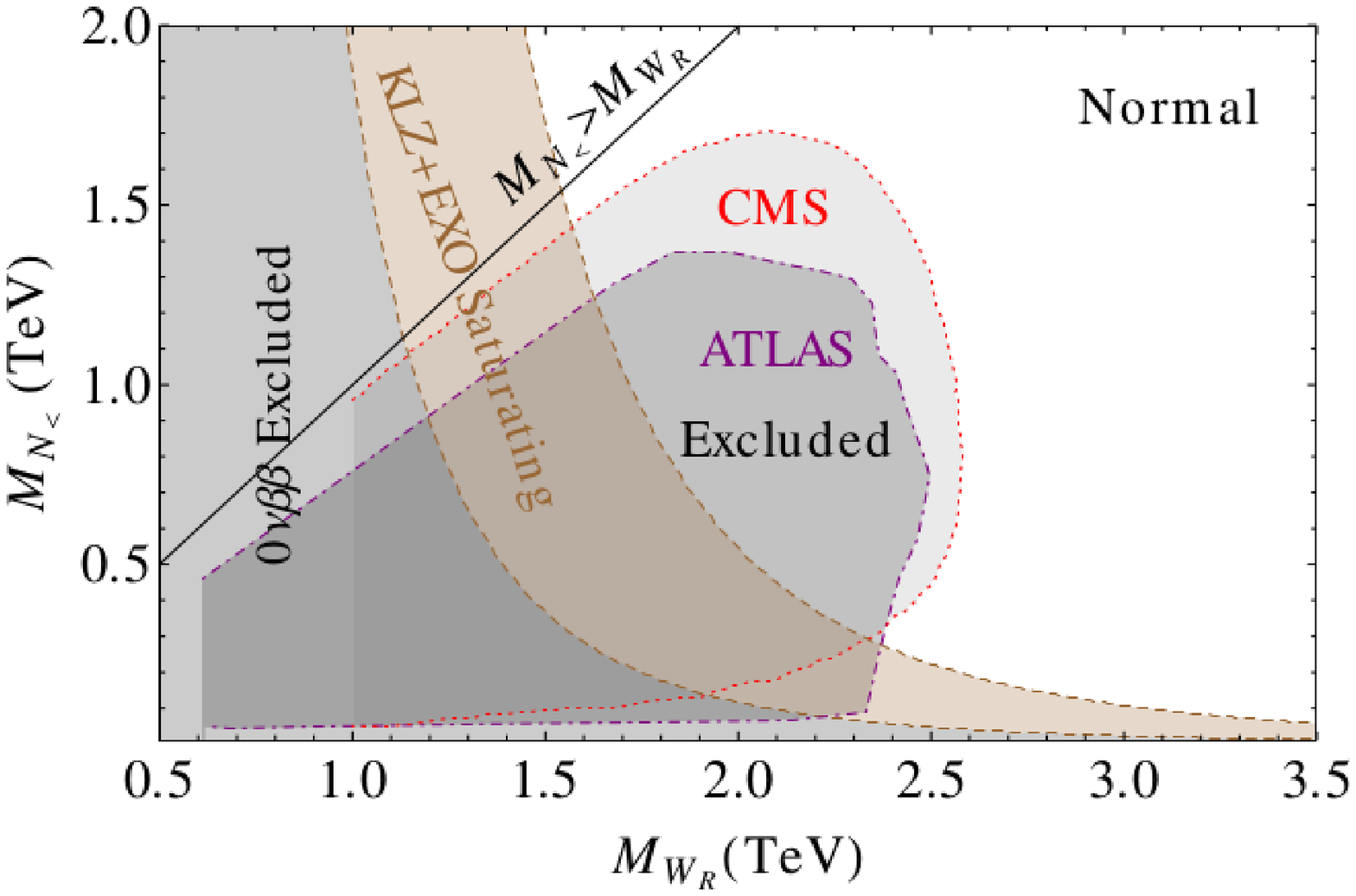}
\caption{\label{fig:LR1}Left: Lifetime of double beta decay if the standard
  and the heavy neutrino exchange with right-handed current diagrams
  are added in type-II dominance. Right: Comparison of double beta
  decay and LHC limits on heavy neutrino masses and right-handed $W_R$
  mass, demonstrating their complementarity  (from \protect\cite{Dev:2013vxa}). }
\end{center}
\end{figure}

Obviously many diagrams can contribute at the same
time and interference between the different diagrams can arise. This nicely demonstrates the
importance of the ideas discussed above to discriminate the
various mechanisms.
Another example for the consequences of several diagrams, 
adding for instance the heavy neutrino
exchange with right-handed currents to the standard amplitude in the case
of type-II dominance is illustrated in Fig.\ \ref{fig:LR1}. A lower limit on
the smallest neutrino mass results, in contrast to the upper limit deduced if
only the standard diagram was taken into account. 

\subsection{$R$--Parity Violating Supersymmetry}
\label{sec:rpv}

The MSSM (minimal supersymmetric extension of the Standard Model) 
assumes the existence of a discrete $Z_2$ symmetry, called $R$-parity.  
This symmetry guarantees the lightest supersymmetric particle 
to be stable, providing a dark matter candidate for cosmology and
avoiding too fast proton decay. Since a convincing theoretical reason
for $R$-parity conservation is lacking, one can investigate the
consequences of its violation. 
Using discrete symmetries one can avoid terms that lead to proton
decay and is left with a superpotential including
the LNV terms  
\begin{equation}\label{R-viol} 
W_{RPV} = 
\lambda'_{ijk}L_i Q_j {\bar D}_k
\,,
\end{equation}
where $i,j, k$ are generation indices. 
Note that the lepton number
violation is by one unit, hence two vertices are required for
$0\nu\beta\beta$, which occurs through long- and short-range Feynman graphs
involving the exchange of superpartners 
\cite{Mohapatra:1986su,Hirsch:1995zi, Hirsch:1995ek, Hirsch:1995cg,
Pas:1998nn}. The short-range contribution has been discussed in
\cite{Mohapatra:1986su,Hirsch:1995zi, Hirsch:1995ek}.  
Combining the half-life limit \cite{Agostini:2013mzu}
with the corrected numerical values \cite{Deppisch:2012nb}  
of the nuclear matrix elements first
calculated in \cite{Hirsch:1995ek} leads to the 
limit on $\lambda_{111}^{'}$ given by
\begin{equation}
	\lambda_{111}^{'}\leq 
		2\cdot 10^{-4}\Big(\frac{m_{\tilde{q}}}{100\, {\rm GeV}} \Big)^2
		\Big(\frac{m_{\tilde{g}}}{100 \,{\rm GeV}} \Big)^{1/2}\,,
\end{equation}
where we have assumed dominance of the gluino exchange diagram and 
took $m_{\tilde{d}_{R}}=m_{\tilde{u}_{L}}\equiv m_{\tilde{q}}$ for the
exchanged squarks. 

\begin{figure}[t]
\begin{center}
\includegraphics[width=.315\textwidth,angle=0]{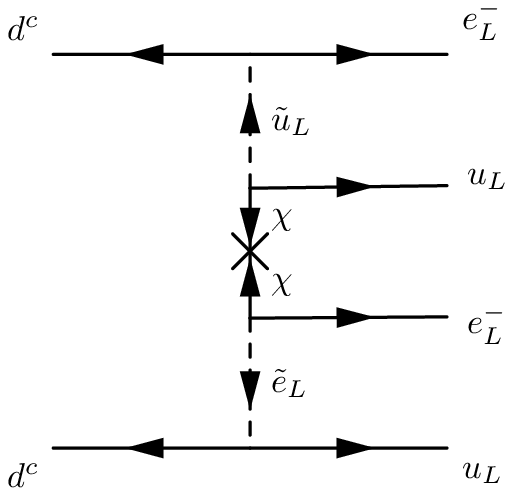}
\includegraphics[width=.315\textwidth,angle=0]{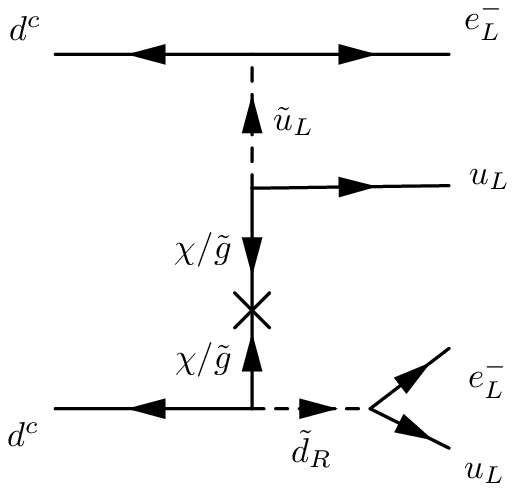}
\includegraphics[width=.315\textwidth,angle=0]{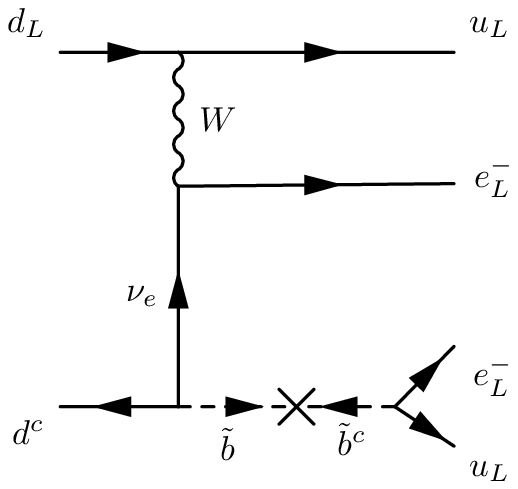}
\caption{\label{fig:RPV}Left and center: exemplary short-range diagrams for  neutrinoless double
  beta decay in $R$-parity violating SUSY. Right: long-range diagram. }
\end{center}
\end{figure}

In addition
$0\nu\beta\beta$ decay is also sensitive to other combinations of the couplings
$\lambda_{ijk}^{'}$. Taking into account the fact that the SUSY
partners of the left- and right-handed quark states can mix with each
other, new diagrams appear in which the neutrino-mediated double beta
decay is accompanied by SUSY exchange in the vertices
\cite{Babu:1995vh, Hirsch:1995cg, Pas:1998nn}, see Fig.\ \ref{fig:RPV}
and note that this is a long-range diagram.  
Assuming the supersymmetric mass parameters of
order 100 GeV, the present GERDA half life limit
implies: $\lambda_{113}^{'} \lambda_{131}^{'}\leq 3 \cdot
10^{-8}$, $\lambda_{112}^{'} \lambda_{121}^{'}\leq 1 \cdot 10^{-6}$. 
Comparable bounds can be deduced from $B$ and $K$ physics which depend however on different superpartner 
masses and are thus complementary to the bounds derived here 
\cite{Allanach:2009xx}.
Recently, the lepton non-universality anomaly at LHCb \cite{Hiller:2014yaa} and the CMS anomaly in the search for right-handed $W$ bosons
have been explained within $R$-parity violating SUSY with 
$\lambda'_{113} = {\cal O}(10^{-3}-10^{-2})$
and  $\lambda'_{112} = {\cal O}(10^{-1})$ and scalar masses in the 
TeV range \cite{Biswas:2014gga}.


\subsection{Leptoquarks}
\label{sec:leptoquarks}

Leptoquarks (LQs) are hypothetical bosons (scalar or vector particles) with  couplings 
to both leptons and quarks which appear for instance in GUTs, 
extended technicolor or compositeness 
models. LQs which conserve baryon number can be possibly within reach of
accelerator experiments \cite{Buchmuller:1986zs}. 
For a detailed list on constraints from non-accelerator 
searches see, for example \cite{Davidson:1993qk} and \cite{Nakamura:2010zzi}. 
The mixing of different LQ multiplets by a possible leptoquark-Higgs
coupling \cite{Hirsch:1996qy} can lead to long-range contributions to
$0\nu\beta\beta$ decay, if these couplings violate lepton number 
\cite{Hirsch:1996ye}. 
%
%
From the lower limit on the $0\nu\beta\beta$ lifetime, bounds on
effective couplings can be derived \cite{Hirsch:1996ye} which are
typically of order 
\begin{equation}
	Y_{LQ-{\rm Higgs}} = {\rm few}\cdot 10^{-6}\,
\end{equation}
for LQ masses of the order of ${\cal O}(200) {\rm GeV}$. 

\subsection{Extra Dimensions}
\label{sec:extradimensions}

Models with more than three space dimensions became popular in recent years
as a way to reduce the four-dimensional Planck scale and 
alleviate this way the hierarchy problem. Extra dimensions have also been suggested as 
a way to generate small Dirac neutrino masses by utilizing the volume
suppressed wave function overlap of a left-handed neutrino confined to
a three-dimensional subspace called the brane and a right-handed neutrino
propagating in the extra-dimensional hyperspace called the bulk
\cite{Dienes:1998sb,ArkaniHamed:1998vp}.
A minimal  higher-dimensional
model implementing  LNV compactifies a 5-dimensional 
theory 
on an $S^1/Z_2$ orbifold, and adds a single
(bulk) sterile neutrino to the field content of  
the SM \cite{Bhattacharyya:2002vf}.   
While the  singlet neutrinos can  freely
propagate   in the bulk, all SM particles are localized
on the $(3+1)$-dimensional brane.

An interesting feature of such 
extra-dimensional models is that the excitations of the sterile neutrino 
in the compactified extra dimensions, a so-called Kaluza-Klein tower of states, 
contribute to the $0\nu\beta\beta$ decay rate. 
The masses of these Kaluza-Klein states are obtained by diagonalizing
the infinitely dimensional Kaluza-Klein mass matrix and result
approximately as
\begin{equation}
  \label{mn}
m_{(n)}\ \approx\  \frac{n}{R}\ + \: \varepsilon\; .
\end{equation}
Here $n$ is the index denoting the Kaluza-Klein excitation, 
$R$ is the radius of the extra
dimension
and $\varepsilon$ is the smallest diagonal entry in the neutrino mass matrix.
As the $m_{(n)}$
range from small masses giving rise to long-range contributions
over the 100~MeV region up to large masses with short-range contributions,
such scenarios constitute a special case which is not included in the
effective operator parametrization described above. 

An important problem of such  extra-dimensional models is  the
generic   prediction  of a   Kaluza-Klein  neutrino  spectrum  with approximately
degenerate masses and opposite  CP parities that leads to an extremely
suppressed contribution to double beta decay and only one $\Delta m^2$
insufficient to explain solar and atmospheric neutrino oscillations. 
If the brane was located at one of the two orbifold fixed points,
the lepton number violating operators thus would be absent as
a consequence  of the $Z_2$ discrete symmetry. If, however, the brane
is shifted  away from the orbifold fixed points, 
the  Kaluza-Klein neutrinos can  couple  to the $W$
bosons with unequal strength, thus avoiding CP-parity
cancellations in the  $0\nu\beta\beta$ amplitude. 
This breaking
of  lepton  number can lead  to   observable effects  in  neutrinoless
double beta decay experiments. 
The size of the brane-shift  can then  be  determined from  the
$0\nu\beta\beta$ lifetime or its upper bound.  

This leads to a nuclear matrix element
depending on the  Kaluza-Klein neutrino masses  $m_{(n)}$,
and thus to predictions for the double beta decay observable 
that depend on
the  double beta  emitter isotope  used in the experiment.   
Another  interesting   property of this model is that the amplitude of
the decay is not bounded from above by the mass eigenvalues of the light neutrinos:
It can  be  close to the experimental limit even for 
an almost vanishing lightest neutrino mass which
constitutes
a  rather  unique property of such extra-dimensional brane-shifted
scenarios.

\section{Lepton Number Violation at Colliders, Double Beta Decay and the 
Baryon Asymmetry of the Universe}
\label{sec:YB}

\begin{figure}[!t]
\centering
\includegraphics[width=3in]{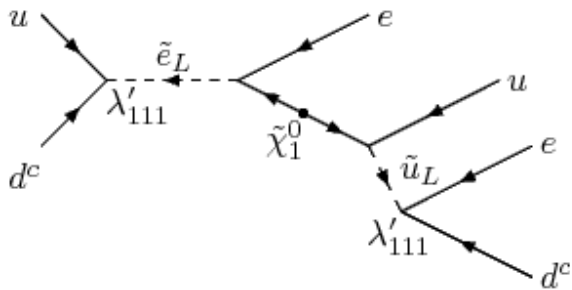}
\caption{Neutrinoless double beta decay at the LHC: the case for $R$-parity violation.  Two quarks in the initial state are
converted into a same-sign di-lepton signal and two jets (from \protect{\cite{Allanach:2009iv}}).} 
\label{rpvlhc}
\end{figure}

In this Section we deal with the links between neutrinoless double beta
decay and lepton number violation processes at colliders and in
cosmology, 
with the latter ones having important consequences for baryogenesis.
As mentioned already in the last Section, while $0\nu\beta\beta$ decay 
provides the best possibility to search 
for light massive Majorana neutrinos, lepton number violation 
as featured in the short-range contributions
can in general be probed also in collider processes.  
For example, as discussed for left-right symmetric models 
\cite{Keung:1983uu,Tello:2010am} (see Fig.\ \ref{fig:LR1}) 
and $R$-parity violating supersymmetry \cite{Allanach:2009iv,Allanach:2009xx}, 
the short-range contribution can easily be crossed into a diagram with two 
quarks in the initial state
where resonant production of a heavy particle leads to a same-sign dilepton signature plus two jets at the LHC, see Fig.~\ref{rpvlhc}. 
If one wants to discuss the LHC bounds in a model-independent way 
it is necessary to specify which particles are propagating in the 
inner legs, which requires a decomposition of the $d=9$ operator in the effective mass approach discussed above. 
Such a decomposition has been worked out in \cite{Bonnet:2012kh} where two different possible 
topologies have been identified. While topology 1 contains two bosons and a
fermion in the internal lines (like the right-handed analogue of the standard diagram),
topology 2 contains an internal 3-boson-vertex (like the triplet
exchange diagram). 
This decomposition was applied to the LHC analogue of $0\nu\beta\beta$ decay and first results for topology 1 
have been worked out in  \cite{Helo:2013dla,Helo:2013ika}. 
The conclusion reached was that with the exception of leptoquark exchange, the 
LHC was typically more sensitive than $0\nu\beta\beta$ decay on the
short-range operators.  Thus one could infer that typically and with
some exceptions
either an observation of  $0\nu\beta\beta$ decay would imply an
LHC signal of LNV as well (in turn, no sign of LNV at the LHC would
exclude an observation of  $0\nu\beta\beta$ decay),  
or $0\nu\beta\beta$ decay would be triggered by a long-range 
mechanism.

In addition, as has been mentioned before, lepton number violation and baryon 
number violation are closely interrelated. More concretely, 
an observation of lepton number violation at low energies has important consequences for a pre-existing lepton asymmetry in the
Universe as the observation of LNV at the LHC 
will yield a lower bound on the washout factor for the lepton asymmetry in the early Universe. 
In \cite{Deppisch:2013jxa} it has thus been pointed out  that any observation of lepton number violation at the LHC will falsify high-scale leptogenesis. 
It is easy to see that
this argument can be
extended even further (for further details see \cite{Deppisch:2015yqa,Harz:2015fwa}).

Just like the combination of $B-L$ violating heavy neutrino 
decays in leptogenesis with $B+L$ violating sphaleron  processes can generate a baryon asymmetry, low energy $B-L$ violation observed at the LHC or elsewhere in combination with $B+L$ violating sphaleron  processes
will wash out any pre-existing baryon asymmetry, whatever of the concrete mechanism of baryogenesis is.

By combining this argument with the results of~\cite{Helo:2013dla,Helo:2013ika} discussed above,  
one can conclude that an observation of short-range $0\nu\beta\beta$
decay will typically imply that LNV processes should be detected at
the LHC as well, and this in turn will falsify standard thermal leptogenesis and in general any high-scale scenario of baryogenesis. 
While the observation that low-energy LNV is dangerous for baryogenesis is not new (see e.g.\ 
\cite{Fukugita:1990gb,Gelmini:1992bz,KlapdorKleingrothaus:1999bd,Frere:2008ct,Hollenberg:2011kq}), only quite recently it has been realized in \cite{Deppisch:2015yqa} that the argument applies for all short range contributions d) and also for the long-range contribution b) in Fig.~\ref{general}. 

It should be stressed of course
that these arguments are rather general and various loopholes exist in specific models:

\begin{itemize}

\item
Scenarios where LNV exists only for (a) specific flavor(s). As $0\nu\beta\beta$ decay probes $\Delta L_e=2$ LNV, only,
it may be possible that lepton number could still be conserved in the $\tau$ flavor which is not necessarily in
equilibrium with the $e$ and $\mu$ flavors in the early Universe \cite{Deppisch:2013jxa}.
It has been discussed in \cite{Deppisch:2015yqa}, however, that an observation of 
lepton flavor violating (LFV) decays such as $\tau \rightarrow \mu \gamma$
may require LFV couplings large enough to wash out such a flavor specific lepton asymmetry when combined with LNV observed in
a different flavor sector; 

\item
Models with hidden sectors, new symmetries and/or conserved charges may protect a baryon asymmetry against 
LNV washout as proposed for the example of
hypercharge by \cite{Antaramian:1993nt}; 

\item
Models where lepton number 
is broken at a scale below the electroweak phase transition where sphalerons are no longer active.

\end{itemize}

As in general an observation of low energy 
LNV would invalidate any high-scale generation 
of the baryon asymmetry though, such
protection mechanisms should be addressed explicitly in any model combining low-scale LNV with high-scale baryogenesis.

\begin{figure}[!t]
\centering
\includegraphics[width=5in]{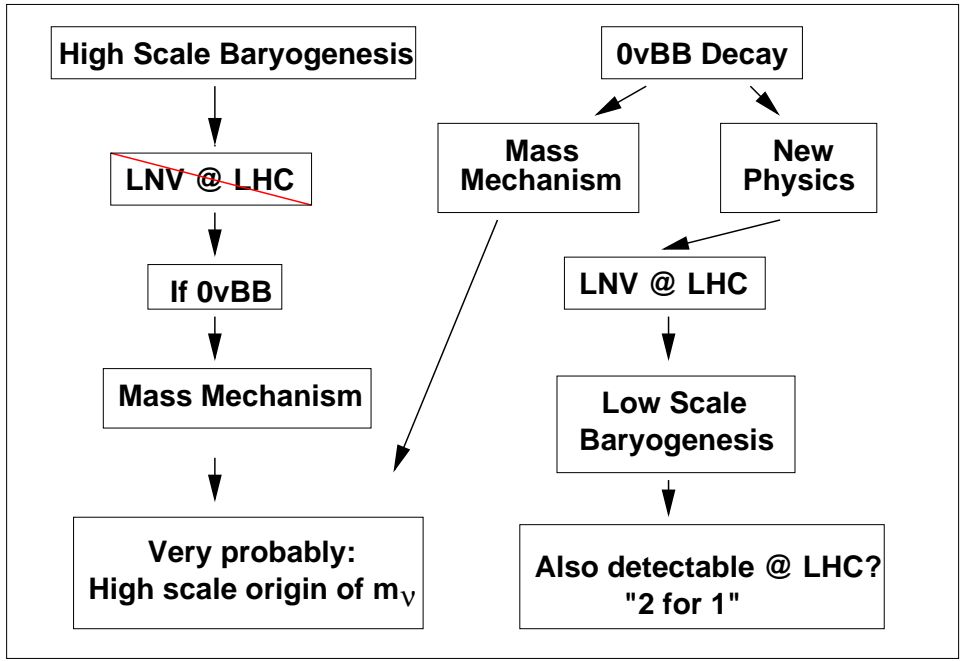}
\caption{The relation of $0\nu\beta\beta$ decay, LNV at the LHC and 
baryogenesis depicted as a logic tree. For details, see text
(from \protect{\cite{Harz:2015fwa}}).
}
\label{Con}
\end{figure}

By building up on the arguments given above, one can conclude, keeping
the above mentioned loopholes in mind, that
if $0\nu\beta\beta$ decay is observed, it is
either triggered by a long-range mechanism, such as the standard interpretation with a light Majorana neutrino 
mass, or due to a short-range mechanism. In this latter case it is very probable that lepton number is observed at the LHC.
This further implies that baryogenesis is a low-scale phenomenon which also may be observable at the LHC or other experiments.

If, on the other hand, the baryon asymmetry is generated at a high scale, LNV will not be observable at the LHC.
If, in this case, $0\nu\beta\beta$ decay will be found, it will typically be triggered by a long-range mechanism. 
In combination with the assumption that we do not have a hint for lepton number violation at a low-scale and that on the
other hand a high scale mechanism is responsible for the generation of the baryon asymmetry, 
this case will probably point towards a high-scale origin of
the neutrino mass as well, such as a type-I seesaw mechanism in combination with leptogenesis.

To summarize this discussion, an observation of $0\nu\beta\beta$ decay 
will  (see Fig.~\ref{Con})
either imply LNV at the LHC and low-scale baryogenesis and thus a possible observation of both processes in the near future,
or very probably point towards a high-scale origin of both neutrino masses and baryogenesis.

\section{Conclusions}

The discovery of lepton number violation would have far-reaching consequences
affecting deeply our thinking about fundamental physics, including  
our ideas about unification and our
understanding of the generation of the baryon asymmetry of the Universe.
Neutrinoless double beta decay and lepton number violation thus
remain fields that enjoy large interest from both experimental and
theoretical communities in nuclear and particle physics. 
In this review we have tried to summarize the multifaceted relations between
neutrinoless double beta decay, neutrino physics 
 and new physics beyond the Standard Model.
The continuous theoretical and
experimental efforts around the world justify the hope that we may not be 
too far away from identifying the origin of lepton number violation. 

\label{sec:conc}

\ack
WR was supported by the Max Planck Society in the project MANITOP.


\section*{References}

\begin{thebibliography}{99}

\bibitem{KlapdorKleingrothaus:2000sn}
  H.~V.~Klapdor-Kleingrothaus, A.~Dietz, L.~Baudis, G.~Heusser, I.~V.~Krivosheina, S.~Kolb, B.~Majorovits and H.~P\"as {\it et al.},
  Eur.\ Phys.\ J.\ A {\bf 12} (2001) 147
  [hep-ph/0103062].

\bibitem{Auger:2012ar}
  M.~Auger {\it et al.}  [EXO Collaboration],
  Phys.\ Rev.\ Lett.\  {\bf 109} (2012) 032505
  [arXiv:1205.5608 [hep-ex]].

\bibitem{Gando:2012zm}
  A.~Gando {\it et al.}  [KamLAND-Zen Collaboration],
  Phys.\ Rev.\ Lett.\  {\bf 110} (2013) 6,  062502
  [arXiv:1211.3863 [hep-ex]].


\bibitem{Agostini:2013mzu}
  M.~Agostini {\it et al.}  [GERDA Collaboration],
  Phys.\ Rev.\ Lett.\  {\bf 111} (2013) 12,  122503
  [arXiv:1307.4720 [nucl-ex]].

\bibitem{Albert:2014awa}
  J.~B.~Albert {\it et al.}  [EXO-200 Collaboration],
  Nature {\bf 510} (2014) 229–234
  [arXiv:1402.6956 [nucl-ex]].




\bibitem{Rodejohann:2011mu}
  W.~Rodejohann,
  Int.\ J.\ Mod.\ Phys.\ E {\bf 20} (2011) 1833
  [arXiv:1106.1334 [hep-ph]].


\bibitem{GomezCadenas:2011it}
  J.~J.~Gomez-Cadenas, J.~Martin-Albo, M.~Mezzetto, F.~Monrabal and M.~Sorel,
  Riv.\ Nuovo Cim.\  {\bf 35} (2012) 29
  [arXiv:1109.5515 [hep-ex]].


\bibitem{Elliott:2012sp}
  S.~R.~Elliott,
  Mod.\ Phys.\ Lett.\ A {\bf 27} (2012) 1230009
  [arXiv:1203.1070 [nucl-ex]].

\bibitem{Bilenky:2012qi}
  S.~M.~Bilenky and C.~Giunti,
  Mod.\ Phys.\ Lett.\ A {\bf 27} (2012) 1230015
  [arXiv:1203.5250 [hep-ph]]; 
arXiv:1411.4791 [hep-ph].


\bibitem{Vergados:2012xy}
  J.~D.~Vergados, H.~Ejiri and F.~Simkovic,
  Rept.\ Prog.\ Phys.\  {\bf 75} (2012) 106301
  [arXiv:1205.0649 [hep-ph]].


\bibitem{Rodejohann:2012xd}
  W.~Rodejohann,
  J.\ Phys.\ G {\bf 39} (2012) 124008
  [arXiv:1206.2560 [hep-ph]].



\bibitem{Deppisch:2012nb}
  F.~F.~Deppisch, M.~Hirsch and H.~P\"as,
  J.\ Phys.\ G {\bf 39} (2012) 124007
  [arXiv:1208.0727 [hep-ph]].

\bibitem{Vogel:2012ja}
  P.~Vogel,
  J.\ Phys.\ G {\bf 39} (2012) 124002
  [arXiv:1208.1992 [nucl-th]].


\bibitem{Schwingenheuer:2012zs}
  B.~Schwingenheuer,
  Annalen Phys.\  {\bf 525} (2013) 269
  [arXiv:1210.7432 [hep-ex]].

\bibitem{Petcov:2013poa}
  S.~T.~Petcov,
  Adv.\ High Energy Phys.\  {\bf 2013} (2013) 852987
  [arXiv:1303.5819 [hep-ph]].


\bibitem{Cremonesi:2013vla}
  O.~Cremonesi and M.~Pavan,
  arXiv:1310.4692 [physics.ins-det].

\bibitem{Haxton:1985am}
  W.~C.~Haxton and G.~J.~Stephenson,
  Prog.\ Part.\ Nucl.\ Phys.\  {\bf 12} (1984) 409.

\bibitem{Doi:1985dx}
  M.~Doi, T.~Kotani and E.~Takasugi,
  Prog.\ Theor.\ Phys.\ Suppl.\  {\bf 83} (1985) 1.


\bibitem{Vergados:2002pv}
  J.~D.~Vergados,
  Phys.\ Rept.\  {\bf 361} (2002) 1
  [hep-ph/0209347].

\bibitem{Avignone:2007fu}
  F.~T.~Avignone, III, S.~R.~Elliott and J.~Engel,
  Rev.\ Mod.\ Phys.\  {\bf 80} (2008) 481
  [arXiv:0708.1033 [nucl-ex]].



\bibitem{Simkovic:2007vu}
  F.~Simkovic, A.~Faessler, V.~Rodin, P.~Vogel and J.~Engel,
  Phys.\ Rev.\ C {\bf 77} (2008) 045503
  [arXiv:0710.2055 [nucl-th]].

\bibitem{Heeck:2013rpa}
  J.~Heeck and W.~Rodejohann,
  Europhys.\ Lett.\  {\bf 103} (2013) 32001
  [arXiv:1306.0580 [hep-ph]].


\bibitem{Osipowicz:2001sq}
  A.~Osipowicz {\it et al.}  [KATRIN Collaboration],
  hep-ex/0109033.


\bibitem{Monreal:2009za}
  B.~Monreal and J.~A.~Formaggio,
  Phys.\ Rev.\ D {\bf 80} (2009) 051301
  [arXiv:0904.2860 [nucl-ex]].



\bibitem{Blaum:2013pfu}
  K.~Blaum, A.~Doerr, C.~E.~Duellmann, K.~Eberhardt, S.~Eliseev, C.~Enss, A.~Faessler and A.~Fleischmann {\it et al.},
  arXiv:1306.2655 [physics.ins-det].


\bibitem{Monfardini:2005dk}
  A.~Monfardini, C.~Arnaboldi, C.~Brofferio, S.~Capelli, F.~Capozzi, O.~Cremonesi, C.~Enss and E.~Fiorini {\it et al.},
  Nucl.\ Instrum.\ Meth.\ A {\bf 559} (2006) 346
  [hep-ex/0509038].


\bibitem{Lesgourgues:2014zoa}
  J.~Lesgourgues and S.~Pastor,
  New J.\ Phys.\  {\bf 16} (2014) 065002
  [arXiv:1404.1740 [hep-ph]].



\bibitem{Akhmedov:2014kxa}
  E.~Akhmedov,
  arXiv:1412.3320 [hep-ph].


\bibitem{Forero:2014bxa}
  D.~V.~Forero, M.~Tortola and J.~W.~F.~Valle,
  Phys.\ Rev.\ D {\bf 90} (2014) 9,  093006
  [arXiv:1405.7540 [hep-ph]].

\bibitem{Capozzi:2013csa}
  F.~Capozzi, G.~L.~Fogli, E.~Lisi, A.~Marrone, D.~Montanino and A.~Palazzo,
  Phys.\ Rev.\ D {\bf 89} (2014) 9,  093018
  [arXiv:1312.2878 [hep-ph]].

\bibitem{Gonzalez-Garcia:2014bfa}
  M.~C.~Gonzalez-Garcia, M.~Maltoni and T.~Schwetz,
  JHEP {\bf 1411} (2014) 052
  [arXiv:1409.5439 [hep-ph]].

\bibitem{Fogli:2004as}
  G.~L.~Fogli, E.~Lisi, A.~Marrone, A.~Melchiorri, A.~Palazzo, P.~Serra and J.~Silk,
  Phys.\ Rev.\ D {\bf 70} (2004) 113003
  [hep-ph/0408045].


\bibitem{Feruglio:2002af}
  F.~Feruglio, A.~Strumia and F.~Vissani,
  Nucl.\ Phys.\ B {\bf 637} (2002) 345
   [Nucl.\ Phys.\ B {\bf 659} (2003) 359]
  [hep-ph/0201291].

\bibitem{Dell'Oro:2015tia}
  S.~Dell'Oro, S.~Marcocci, M.~Viel and F.~Vissani,
  arXiv:1505.02722 [hep-ph].


\bibitem{Pascoli:2002xq}
  S.~Pascoli and S.~T.~Petcov,
  Phys.\ Lett.\ B {\bf 544} (2002) 239
  [hep-ph/0205022].


\bibitem{Dueck:2011hu}
  A.~Dueck, W.~Rodejohann and K.~Zuber,
  Phys.\ Rev.\ D {\bf 83} (2011) 113010
  [arXiv:1103.4152 [hep-ph]].



\bibitem{Aalseth:2002rf}
  C.~E.~Aalseth {\it et al.}  [IGEX Collaboration],
  Phys.\ Rev.\ D {\bf 65} (2002) 092007
  [hep-ex/0202026].

\bibitem{Kotila:2012zza}
  J.~Kotila and F.~Iachello,
  Phys.\ Rev.\ C {\bf 85} (2012) 034316
  [arXiv:1209.5722 [nucl-th]].

\bibitem{Mirea:2014dza}
  M.~Mirea, T.~Pahomi and S.~Stoica,
  arXiv:1411.5506 [nucl-th].

\bibitem{Dev:2013vxa}
  P.~S.~Bhupal Dev, S.~Goswami, M.~Mitra and W.~Rodejohann,
  Phys.\ Rev.\ D {\bf 88} (2013) 091301
  [arXiv:1305.0056 [hep-ph]].



\bibitem{Guzowski:2015saa}
  P.~Guzowski, L.~Barnes, J.~Evans, G.~Karagiorgi, N.~McCabe and S.~Soldner-Rembold,
  arXiv:1504.03600 [hep-ex].

\bibitem{Rodriguez:2010mn} 
  T.~R.~Rodriguez and G.~Martinez-Pinedo,
  Phys.\ Rev.\ Lett.\  {\bf 105} (2010) 252503.

\bibitem{Menendez:2008jp} 
  J.~Menendez, A.~Poves, E.~Caurier and F.~Nowacki,
  Nucl.\ Phys.\ A {\bf 818} (2009) 139.

\bibitem{Barea:2013bz} 
  J.~Barea, J.~Kotila and F.~Iachello,
  Phys.\ Rev.\ C {\bf 87} (2013) 014315. 

\bibitem{Suhonen:2010zzc} 
  J.~Suhonen and O.~Civitarese,
  Nucl.\ Phys.\ A {\bf 847} (2010) 207.

  

\bibitem{Meroni:2012qf} 
  A.~Meroni, S.~T.~Petcov and F.~Simkovic,
  JHEP {\bf 1302} (2013) 025.


\bibitem{Simkovic:2013qiy} 
  F.~Simkovic, V.~Rodin, A.~Faessler and P.~Vogel,
  Phys.\ Rev.\ C {\bf 87} (2013) 045501.

\bibitem{Mustonen:2013zu} 
  M.~T.~Mustonen and J.~Engel,
Phys.\ Rev.\ C {\bf 87} (2013) 6,  064302
  [arXiv:1301.6997 [nucl-th]].



\bibitem{Altarelli:2010gt}
  G.~Altarelli and F.~Feruglio,
  Rev.\ Mod.\ Phys.\  {\bf 82} (2010) 2701
  [arXiv:1002.0211 [hep-ph]].



\bibitem{Ishimori:2010au}
  H.~Ishimori, T.~Kobayashi, H.~Ohki, Y.~Shimizu, H.~Okada and M.~Tanimoto,
  Prog.\ Theor.\ Phys.\ Suppl.\  {\bf 183} (2010) 1
  [arXiv:1003.3552 [hep-th]].



\bibitem{Barry:2010yk}
  J.~Barry and W.~Rodejohann,
  Nucl.\ Phys.\ B {\bf 842} (2011) 33
  [arXiv:1007.5217 [hep-ph]].


\bibitem{Dorame:2011eb}
  L.~Dorame, D.~Meloni, S.~Morisi, E.~Peinado and J.~W.~F.~Valle,
  Nucl.\ Phys.\ B {\bf 861} (2012) 259
  [arXiv:1111.5614 [hep-ph]].


\bibitem{Dorame:2012zv}
  L.~Dorame, S.~Morisi, E.~Peinado, J.~W.~F.~Valle and A.~D.~Rojas,
  Phys.\ Rev.\ D {\bf 86} (2012) 056001
  [arXiv:1203.0155 [hep-ph]].


\bibitem{King:2013psa}
  S.~F.~King, A.~Merle and A.~J.~Stuart,
  JHEP {\bf 1312} (2013) 005
  [arXiv:1307.2901 [hep-ph]].


\bibitem{Dueck:2013gca}
  A.~Dueck and W.~Rodejohann,
  JHEP {\bf 1309} (2013) 024
  [arXiv:1306.4468 [hep-ph]].



\bibitem{Abazajian:2012ys}
  K.~N.~Abazajian, M.~A.~Acero, S.~K.~Agarwalla, A.~A.~Aguilar-Arevalo, C.~H.~Albright, S.~Antusch, C.~A.~Arguelles and A.~B.~Balantekin {\it et al.},
  arXiv:1204.5379 [hep-ph].





\bibitem{Barry:2011wb}
  J.~Barry, W.~Rodejohann and H.~Zhang,
  JHEP {\bf 1107} (2011) 091
  [arXiv:1105.3911 [hep-ph]].


\bibitem{Giunti:2012tn}
  C.~Giunti, M.~Laveder, Y.~F.~Li, Q.~Y.~Liu and H.~W.~Long,
  Phys.\ Rev.\ D {\bf 86} (2012) 113014
  [arXiv:1210.5715 [hep-ph]].

\bibitem{Girardi:2013zra}
  I.~Girardi, A.~Meroni and S.~T.~Petcov,
  JHEP {\bf 1311} (2013) 146
  [arXiv:1308.5802 [hep-ph]].
  
\bibitem{Giunti:2015kza}
  C.~Giunti and E.~M.~Zavanin,
  arXiv:1505.00978 [hep-ph].






 


\bibitem{Schechter:1981bd}
  J.~Schechter and J.~W.~F.~Valle,
  Phys.\ Rev.\ D {\bf 25} (1982) 2951.

\bibitem{Nieves:1984sn}
  J.~F.~Nieves,
  Phys.\ Lett.\ B {\bf 147} (1984) 375.

\bibitem{Takasugi:1984xr}
  E.~Takasugi,
  Phys.\ Lett.\ B {\bf 149} (1984) 372.

\bibitem{Hirsch:1998mc}
  M.~Hirsch, H.~V.~Klapdor-Kleingrothaus and S.~G.~Kovalenko,
  Nucl.\ Phys.\ Proc.\ Suppl.\  {\bf 62} (1998) 224.

\bibitem{Hirsch:2006yk}
  M.~Hirsch, S.~Kovalenko and I.~Schmidt,
  Phys.\ Lett.\ B {\bf 642} (2006) 106
  [hep-ph/0608207].

\bibitem{Duerr:2011zd}
  M.~Duerr, M.~Lindner and A.~Merle,
  JHEP {\bf 1106} (2011) 091
  [arXiv:1105.0901 [hep-ph]].

\bibitem{Helo:2015fba}
  J.~C.~Helo, M.~Hirsch, T.~Ota and F.~A.~P.~d.~Santos,
  JHEP {\bf 1505} (2015) 092
  [arXiv:1502.05188 [hep-ph]].

\bibitem{Pas:1999fc}
  H.~P\"as, M.~Hirsch, H.~V.~Klapdor-Kleingrothaus and S.~G.~Kovalenko,
  Phys.\ Lett.\ B {\bf 453} (1999) 194.


\bibitem{Pas:2000vn}
  H.~P\"as, M.~Hirsch, H.~V.~Klapdor-Kleingrothaus and S.~G.~Kovalenko,
  Phys.\ Lett.\ B {\bf 498} (2001) 35
  [hep-ph/0008182].

  
\bibitem{Deppisch:2006hb}
  F.~Deppisch and H.~P\"as,
  Phys.\ Rev.\ Lett.\  {\bf 98} (2007) 232501
  [hep-ph/0612165].


\bibitem{Gehman:2007qg}
  V.~M.~Gehman and S.~R.~Elliott,
  J.\ Phys.\ G {\bf 34}, 667 (2007)
  [Erratum-ibid.\ G {\bf 35}, 029701 (2008)].

\bibitem{Faessler:2011qw}
  A.~Faessler, A.~Meroni, S.~T.~Petcov, F.~Simkovic and J.~Vergados,
  Phys.\ Rev.\ D {\bf 83} (2011) 113003
  [arXiv:1103.2434 [hep-ph]].


  
\bibitem{Arnold:2010tu}
  R.~Arnold {\it et al.} [SuperNEMO Collaboration],
  Eur.\ Phys.\ J.\ C {\bf 70} (2010) 927.
  

\bibitem{Helo:2013dla}
  J.~C.~Helo, M.~Hirsch, S.~G.~Kovalenko and H.~P\"as,
  Phys.\ Rev.\ D {\bf 88} (2013) 1,  011901
  [arXiv:1303.0899 [hep-ph]].
  
\bibitem{Helo:2013ika}
  J.~C.~Helo, M.~Hirsch, H.~P\"as and S.~G.~Kovalenko,
  Phys.\ Rev.\ D {\bf 88} (2013) 073011
  [arXiv:1307.4849 [hep-ph]].



\bibitem{Muto:1989cd}
K.~Muto, E.~Bender and H.~Klapdor,
\newblock Z.Phys. {\bf A334}, 187 (1989).

\bibitem{Hirsch:1996qw}
M.~Hirsch, H.~Klapdor-Kleingrothaus and O.~Panella,
\newblock Phys.Lett. {\bf B374}, 7 (1996), [hep-ph/9602306].


\bibitem{Barry:2013xxa}
  J.~Barry and W.~Rodejohann,
  JHEP {\bf 1309} (2013) 153
  [arXiv:1303.6324 [hep-ph]].

\bibitem{Dev:2014xea}
  P.~S.~B.~Dev, S.~Goswami and M.~Mitra,
  Phys.\ Rev.\ D {\bf 91} (2015) 11,  113004
  [arXiv:1405.1399 [hep-ph]].


\bibitem{Ibarra:2010xw}
  A.~Ibarra, E.~Molinaro and S.~T.~Petcov,
  JHEP {\bf 1009} (2010) 108
  [arXiv:1007.2378 [hep-ph]].

\bibitem{Mitra:2011qr}
  M.~Mitra, G.~Senjanovic and F.~Vissani,
  Nucl.\ Phys.\ B {\bf 856} (2012) 26
  [arXiv:1108.0004 [hep-ph]].

\bibitem{Tello:2010am}
  V.~Tello, M.~Nemevsek, F.~Nesti, G.~Senjanovic and F.~Vissani,
  Phys.\ Rev.\ Lett.\  {\bf 106} (2011) 151801
  [arXiv:1011.3522 [hep-ph]].

\bibitem{Mohapatra:1986su}
  R.~N.~Mohapatra,
  Phys.\ Rev.\ D {\bf 34} (1986) 3457.

\bibitem{Hirsch:1995zi}
M.~Hirsch, H.~Klapdor-Kleingrothaus and S.~Kovalenko,
\newblock Phys.Rev.Lett. {\bf 75}, 17 (1995).

\bibitem{Hirsch:1995ek}
M.~Hirsch, H.~Klapdor-Kleingrothaus and S.~Kovalenko,
\newblock Phys.Rev. {\bf D53}, 1329 (1996), [hep-ph/9502385].



\bibitem{Hirsch:1995cg}
M.~Hirsch, H.~Klapdor-Kleingrothaus and S.~Kovalenko,
\newblock Phys.Lett. {\bf B372}, 181 (1996), [hep-ph/9512237].

\bibitem{Pas:1998nn}
H.~P{\"a}s, M.~Hirsch and H.~Klapdor-Kleingrothaus,
\newblock Phys.Lett. {\bf B459}, 450 (1999), [hep-ph/9810382].

\bibitem{Babu:1995vh}
K.~Babu and R.~Mohapatra,
\newblock Phys.Rev.Lett. {\bf 75}, 2276 (1995), [hep-ph/9506354].

\bibitem{Allanach:2009xx}
  B.~C.~Allanach, C.~H.~Kom and H.~P\"as,
  JHEP {\bf 0910} (2009) 026
  [arXiv:0903.0347 [hep-ph]].

\bibitem{Hiller:2014yaa}
  G.~Hiller and M.~Schmaltz,
  Phys.\ Rev.\ D {\bf 90} (2014) 054014
  [arXiv:1408.1627 [hep-ph]].
  
\bibitem{Biswas:2014gga}
  S.~Biswas, D.~Chowdhury, S.~Han and S.~J.~Lee,
  JHEP {\bf 1502} (2015) 142
  [arXiv:1409.0882 [hep-ph]].


\bibitem{Buchmuller:1986zs}
W.~Buchm{\"u}ller, R.~R{\"u}ckl and D.~Wyler,
\newblock Phys.Lett. {\bf B191}, 442 (1987).

\bibitem{Davidson:1993qk}
S.~Davidson, D.~C. Bailey and B.~A. Campbell,
\newblock Z.Phys. {\bf C61}, 613 (1994), [hep-ph/9309310].

\bibitem{Nakamura:2010zzi}
Particle Data Group, K.~Nakamura {\em et~al.},
\newblock J.Phys.G {\bf G37}, 075021 (2010).

\bibitem{Hirsch:1996qy}
M.~Hirsch, H.~Klapdor-Kleingrothaus and S.~Kovalenko,
\newblock Phys.Lett. {\bf B378}, 17 (1996), [hep-ph/9602305].

\bibitem{Hirsch:1996ye}
M.~Hirsch, H.~Klapdor-Kleingrothaus and S.~Kovalenko,
\newblock Phys.Rev. {\bf D54}, 4207 (1996), [hep-ph/9603213].





\bibitem{Dienes:1998sb}
  K.~R.~Dienes, E.~Dudas and T.~Gherghetta,
  Nucl.\ Phys.\ B {\bf 557} (1999) 25
  [hep-ph/9811428].


\bibitem{ArkaniHamed:1998vp}
  N.~Arkani-Hamed, S.~Dimopoulos, G.~R.~Dvali and J.~March-Russell,
  Phys.\ Rev.\ D {\bf 65} (2002) 024032
  [hep-ph/9811448].


\bibitem{Bhattacharyya:2002vf}
G.~Bhattacharyya, H.~Klapdor-Kleingrothaus, H.~P{\"a}s and A.~Pilaftsis,
\newblock Phys.Rev. {\bf D67}, 113001 (2003), [hep-ph/0212169].
    
  
 

\bibitem{Keung:1983uu}
  W.~Y.~Keung and G.~Senjanovic,
  Phys.\ Rev.\ Lett.\  {\bf 50} (1983) 1427.
  

 
\bibitem{Allanach:2009iv}
  B.~C.~Allanach, C.~H.~Kom and H.~P\"as,
  Phys.\ Rev.\ Lett.\  {\bf 103} (2009) 091801
  [arXiv:0902.4697 [hep-ph]].



\bibitem{Bonnet:2012kh}
  F.~Bonnet, M.~Hirsch, T.~Ota and W.~Winter,
  JHEP {\bf 1303} (2013) 055
   [JHEP {\bf 1404} (2014) 090]
  [arXiv:1212.3045 [hep-ph]].
  
\bibitem{Deppisch:2013jxa}
  F.~F.~Deppisch, J.~Harz and M.~Hirsch,
  Phys.\ Rev.\ Lett.\  {\bf 112} (2014) 221601
  [arXiv:1312.4447 [hep-ph]].


\bibitem{Harz:2015fwa}
  J.~Harz, W.~C.~Huang and H.~P\"as,
  arXiv:1505.07632 [hep-ph].


\bibitem{Deppisch:2015yqa}
  F.~F.~Deppisch, J.~Harz, M.~Hirsch, W.~C.~Huang and H.~P\"as,
  arXiv:1503.04825 [hep-ph].
  

\bibitem{Fukugita:1990gb}
  M.~Fukugita and T.~Yanagida,
  Phys.\ Rev.\ D {\bf 42} (1990) 1285.
  
\bibitem{Gelmini:1992bz} 
  G.~Gelmini and T.~Yanagida,
  Phys.\ Lett.\ B {\bf 294}, 53 (1992).
  
\bibitem{KlapdorKleingrothaus:1999bd}
  H.~V.~Klapdor-Kleingrothaus, S.~Kolb and V.~A.~Kuzmin,
  Phys.\ Rev.\ D {\bf 62} (2000) 035014
  [hep-ph/9909546].
  

\bibitem{Frere:2008ct}
  J.~M.~Frere, T.~Hambye and G.~Vertongen,
  JHEP {\bf 0901} (2009) 051
  [arXiv:0806.0841 [hep-ph]].


\bibitem{Hollenberg:2011kq}
  S.~Hollenberg, H.~P\"as and D.~Schalla,
  arXiv:1110.0948 [hep-ph].
  
  
 
\bibitem{Antaramian:1993nt}
  A.~Antaramian, L.~J.~Hall and A.~Rasin,
  Phys.\ Rev.\ D {\bf 49} (1994) 3881
  [hep-ph/9311279].
  


\end{thebibliography}
\end{document}